\documentclass[12pt, preprint]{aastex}

\usepackage{ulem}  

\newcommand\spitzer{\textit{Spitzer}}

\newcommand{\ltsimeq}{\raisebox{-0.6ex}{$\,\stackrel
        {\raisebox{-.2ex}{$\textstyle <$}}{\sim}\,$}}
\newcommand{\gtsimeq}{\raisebox{-0.6ex}{$\,\stackrel
        {\raisebox{-.2ex}{$\textstyle >$}}{\sim}\,$}}

\slugcomment{AJ rev 24.Feb.09}

\received{}
\revised{}
\accepted{}

\shorttitle{Asteroid Distributions in the Ecliptic}
\shortauthors{Ryan, Woodward, et al.}

\begin{document}


\title{THE ASTEROID DISTRIBUTION IN THE ECLIPTIC}


\author{
ERIN LEE RYAN\altaffilmark{1}, 
CHARLES E. WOODWARD\altaffilmark{1}, 
ANDREA DIPAOLO\altaffilmark{2}, 
JACOPO FARINATO\altaffilmark{3}, 
EMANUELE GIALLONGO\altaffilmark{2}, 
ROLAND GREDEL\altaffilmark{4}, 
JOHN HILL\altaffilmark{5}, 
FERNANDO PEDICHINI\altaffilmark{2}, 
RICHARD POGGE\altaffilmark{6}, 
ROBERTO RAGAZZONI\altaffilmark{3}  
}


\altaffiltext{1}{Department of Astronomy, School of Physics and 
Astronomy, 116 Church Street, S.~E., University of Minnesota, 
Minneapolis, MN 55455,\ \it{ryan@astro.umn.edu, chelsea@astro.umn.edu}} 

\altaffiltext{2}{Osservatorio Astronomico di Roma, via di Frascati 33, 
I-00040 Monteporzio, Italy}

\altaffiltext{3}{Osservatorio Astronomico di Padova, vicolo 
dell'Osservatorio 5, I-35122 Padova, Italy}

\altaffiltext{4}{Max-Planck-Institut fuer Astronomie, Koenigstuhl 17, 
D-69117 Heidelberg, Germany}

\altaffiltext{5}{Large Binocular Telescope Observatory, University of 
Arizona, 933 N Cherry Ave, Tucson, AZ, 85721-0065}

\altaffiltext{6}{Department of Astronomy, The Ohio State University, 
140 W. 18th Avenue, Columbus, OH 43210-1173}

\begin{abstract}
We present analysis of the asteroid surface density distribution
of main belt asteroids (mean perihelion $\Delta \simeq 2.404$~AU) in five 
ecliptic latitude fields, $-17 \gtsimeq \beta(\degr) \ltsimeq +15$, 
derived from deep \textit{Large Binocular Telescope} (LBT) 
$V-$band (85\% completeness limit $V = 21.3$~mag) and 
\textit{Spitzer Space Telescope} IRAC~8.0~\micron\ (80\% completeness limit 
$\sim 103~\mu$Jy) fields enabling us to probe the 0.5--1.0~km diameter 
asteroid population. We discovered 58 new asteroids in the optical 
survey as well as 41 new bodies in the \textit{Spitzer} fields. 
The derived power law slopes of the number of asteroids per square
degree are similar within each $\sim 5$\degr{} ecliptic latitude 
bin with a mean value of $ -0.111 \pm 0.077$. For the 23 
known asteroids detected in all four IRAC channels mean albedos range 
from $0.24 \pm 0.07$ to $0.10 \pm 0.05$. No low albedo
asteroids ($p_{V}$ $\ltsimeq$ 0.1) were detected in the \textit{Spitzer}
FLS fields, whereas in the SWIRE fields they are frequent. 
The SWIRE data clearly samples asteroids in the middle and outer belts 
providing the first estimates of these km-sized asteroids' albedos. Our 
observed asteroid number densities at optical wavelengths are 
generally consistent with those derived from the Standard Asteroid 
Model within the ecliptic plane. However, we find an over density at 
$\beta \gtsimeq 5$\degr{} in our optical fields, while the infrared 
number densities are under dense by factors of 2 to 3 at all 
ecliptic latitudes.

\end{abstract}



\section{INTRODUCTION\label{intro}}

The present main belt asteroid size distribution provides 
important constraints on models of the original size distribution of 
planetesimals and their collisional evolution. The size-frequency 
distribution, main belt asteroid scale height, and albedo variations 
extant in a large, well-sampled population 
are also essential to model the size distribution of the Near Earth 
Asteroid (NEA) population and to determine the evolution of main belt 
objects as compared to those bodies injected into Earth crossing 
orbits \citep[e.g.,][]{trilling08}. 

Most main belt asteroids are found between 2.2 and 3.4 AU from the sun 
and at ecliptic latitudes nominally less than 20\degr. The size-frequency 
distribution of main belt asteroids with diameters $< 1$~km is 
essentially unknown due to observational flux limits. For asteroids with 
diameters $> 1$~km, the derived size-frequency distribution estimates 
range from $7 \times 10^{5}$~km-sized objects \citep[optical Sloan 
Digital Sky Survey, SDSS;][]{aba04,ivez01} to $(1.2 \pm 0.5) \times 
10^{6}$~km-sized objects \citep[derived from surveys with the
\textit{Infrared Space Observatory}, ISO;][]{tedsd02} within the main 
belt. Recently, \citet{stapelf06} 
have derived a size-frequency estimate for the $> 1$~km-sized bodies for 
two fields at $0\degr$ and $10\degr$ ecliptic latitude from 24~\micron \ 
photometry obtained with the NASA \textit{Spitzer Space Telescope} 
\citep[\spitzer,][]{gehrz07, werner04} that concurs with values derived 
from analysis of the Sloan fields. While the infrared (IR) and optical 
methodologies yield comparable size estimates, these surveys are 
potentially biased as they are unable to detect (either through intrinsic 
sensitivity limits or survey design) fainter objects within the main 
belt. Thus, the size-frequency distribution may not be complete due to 
the preferential selection of a cross section of the asteroid population 
with high albedo values or large sizes. 

\citet{botk05} have modeled the collisional evolution of the size 
distribution of the main belt using the current size-frequency 
distributions within the main belt as well as the size-frequency 
distribution of NEAs to constrain the primordial asteroid distribution. 
Their best-fit models, based on the current size-frequency distribution, 
predict $\simeq 10^{6}$ asteroids within the 0.3 to 1 km size range. 
Collisional models by \citet{obgr05} predict $\simeq 10^{5}$ bodies in a 
similar size range within the main belt. However, the validity of the 
population extrapolation is based on an accurate understanding of the 
current size-frequency distribution necessitating deep observations to 
provide accurate number counts at both optical wavelengths (higher albedo
objects) and IR (larger dark asteroids) wavelengths. The collisional evolution 
model by \citet{botk05} suggests that the primordial main belt contained 
$\sim150$ to 250 times the current population of objects with diameters, 
$D \ltsimeq 1000$~km. The \citet{botk05} model produces a parent 
population of similar magnitude to that required by interpretation of the 
meteoritic record. For instance \citet{wethr89} argues that the surface 
density within the main belt must be at least 100 times higher than that 
currently observed to be consistent with the meteoritic record. The total 
mass of known asteroids in the main belt is 
$5 \times 10^{-4}~M_{\oplus}$. However, both the meteoritic record and 
collisional models require a primordial main belt mass of $7.5 \times 
10^{-2}~M_{\oplus}$ to $1.25 \times 10^{-1}~M_{\oplus}$. The significant 
discrepancy between the mass of the primordial main belt and the current 
mass, primarily derived from large ($D \ge 10$~km) sized asteroids, may 
be resolved by extending the asteroid number census to properly encompass 
smaller sizes while including populations found at higher ecliptic 
latitudes. 

Here we present new observational results to constrain the size-frequency 
distribution of main belt asteroids and their ecliptic scale-height 
derived from a deep optical survey of select ecliptic fields obtained 
with the \textit{Large Binocular Telescope} (LBT) as well as near- 
and mid-IR fields observed with \textit{Spitzer} drawn from the data 
archive. Section~\ref{obs} describes our observations and reduction techniques, 
\S\ref{disc} discusses the newly discovered asteroids and our derivation 
albedo, size-frequency distribution, and scale-height, while 
\S\ref{concl} summarizes our conclusions. 

\section{OBSERVATIONS \& ARCHIVAL ANALYSIS\label{obs}}

\subsection{LBT Observations\label{lbt-ord}}

Optical observations were obtained at the Large Binocular Telescope 
\citep[LBT;][]{hill06} facility of the Mt. Graham International 
Observatory with the blue channel of the Large Binocular Camera 
\citep[LBC;][]{ragaz06, gial07} and a single 8.4~m mirror on various 
nights during 2007 January 16 through 24~UT as part of Science 
Demonstration Time (SDT) activities. The LBC is a wide-field imager 
incorporating four $2048 \times 4608$ pixel CCD detectors with a 
$23\arcmin \times 23\arcmin$ field of view (FOV) and a 0.23\arcsec \ per 
pixel platescale. Four fields at ecliptic latitudes of 0\degr, 5\degr, 
10\degr,\ and 15\degr\ were observed using a series of 4~min exposures in 
the V-band ($\lambda_{o} = 0.55$~\micron; $\Delta\lambda = 
0.094$~\micron) under non photometric conditions with seeing between $\simeq 
1.3$\arcsec $ - \, 3.0$\arcsec. Fields were selected to be at solar 
elongations near 160\degr, such that sufficient asteroid motion on the 
plane of the sky could be detected within the $\sim$~3~hr observational 
baseline. Observations of each field consisted of 2 pointings, with one 
pointing offset by $\sim 7.85$\arcmin, resulting in a total areal 
coverage per field of 0.2835 square degrees. Table~\ref{table:lbtobs} 
provides complete observational details. 

The data were reduced using standard IRAF\footnote{IRAF is distributed by 
the National Optical Astronomy Observatories, which are operated by the 
Association of Universities for Research in Astronomy, Inc., under 
cooperative agreement with the National Science Foundation.} routines in 
the \textit{mscred} package. Images were trimmed, bias-subtracted, and 
flat-fielded using median sky flats created with twilight flats and 
additional data obtained during 2007 January SDT time. Astrometric 
solutions for each data frame were generated using IRAF routines, in 
conjunction with the stellar positions obtained from the USNO A2.0 
astrometric catalog \citep{zach04}. The solutions had a positional accuracy of 
$0.25\arcsec \pm 0.08\arcsec$ for each field and the relative position of 
detected asteroids are referenced to this grid. 

Asteroid detection was performed using a three color method. Images with 
astrometric solutions were displayed in image display tool DS9 
\citep{jm03}, with each individual epoch (Table~\ref{table:lbtobs}) 
loaded into a different color table. Asteroids were identified by pairs 
or triplets of individually colored sources.  From these identifications, 
absolute pixel coordinates for each target were obtained, which were 
subsequently used for photometric measurements. Due to the conditions 
under which the data was obtained, a high signal to noise of 10 is 
required for any asteroid detection. Secondary confirmation of 
this asteroid detection technique was performed by registering individual 
epoch frames by the world coordinate system (WCS) and then subtracting, 
leaving positive and negative asteroid pairs. While this latter technique 
would be preferred over the three color technique, the seeing varied by 
1.3\arcsec\ to 3.0\arcsec\ over each night causing the WCS subtraction 
technique to be less reliable. 

Photometry was performed using a circular aperture with a radius of 10 
pixels for each asteroid coordinate. The photometry was calibrated from 
stellar sources in the fields using published $V-$band magnitudes taken 
from the Tycho and USNO A2.0/YB6 catalogs (mean photometric errors of 
the order $\ltsimeq$ 0.1~mag). Forty stars (10 stars per chip) 
were identified by their astrometric positions and were used to find a 
photometric offset between the measured and reported absolute values from 
Tycho/USNO A2.0. This zero-point photometric offset, which includes the 
mean photometric error, was then used to calibrate the asteroid photometry 
in each pointing as described in Table~\ref{table:lbtobs}, column [7]. 

We detected 62 asteroids in the LBC $V-$band images, of which only four 
are previously known objects with orbital determinations.
The 58 newly discovered asteroids detected in this survey were observed 
in either two or three epochs. Due to the field overlap between 
pointings, some asteroids were detected six times, allowing for both 
precise astrometry and photometry. Table~\ref{table:as_numsum} summarizes 
the number of asteroids found in each field as well as the extrapolated 
number counts per square degree at each latitude and the limiting 
magnitudes for each field.  The extrapolated optical number counts were 
obtained by multiplying the asteroid number counts in a field by the 
number 0.2835 square degree fields necessary to tile a 1 square degree 
field assuming a uniform asteroid density at each latitude. 

\subsection{\textit{Spitzer} Archival Fields\label{saf}}

Asteroid photometry and number counts in select ecliptic fields discussed 
in \S\ref{disc} were derived from new ground-based optical observations 
and point-sources extractions obtained from fields observed with the 
Infrared Camera \citep[IRAC;][]{fazio04} and the Multiband 
Infrared Photometer for \textit{Spitzer} \citep[MIPS;][]{rieke04} as part 
of various IR survey programs retrieved in the \spitzer{} public 
archive. 

Post-pipeline (pipeline v15) basic calibrated data (BCDs) of selected 
\textit{Spitzer} IRAC fields were downloaded from the public archive and 
utilized for asteroid detection. Fields were selected from the ecliptic 
plane component of the \textit{Spitzer} First Look Survey 
\citep[FLS, program identification (PID) 98;][]{vmeadows04}, and 
SWIRE XMM-LSS fields \citep[PID 181;][]{cjl03}. 
IRAC 8.0~\micron \ data is preferable for asteroid detection due to the 
detector platescale ($\simeq 1.2$\arcsec) and detector sensitivity. The 
5$\sigma$ detection limit for a 500 sec IRAC 8.0~\micron \ observation is 
27~$\mu$Jy, enabling detection of small main belt asteroids with high 
signal-to-noise ratios (SNR) in short integration times. For example, the 
8.0~\micron \ flux from a 1~km diameter main belt asteroid radiating as a 
blackbody with an orbital semi-major axis of 2.5 AU (3.2 AU) viewed at 
opposition is $\approx 1096~\mu$Jy ($\approx 20.1~\mu$Jy). Once an 
asteroid candidate is identified at 8.0~\micron \, its sky coordinate can 
then be used to examine pixels at the same location in the IRAC 3.6, 4.5, 
and 5.8~\micron \ images. Although the flux densities for small-sized 
asteroids near the outer belt edge are not easily detectable with IRAC in 
single frames at shorter wavelengths, their motion will separate them 
from the confusion caused by faint extragalactic sources or point spread 
function (PSF) smearing of the telescope. 

Due to limits on observation durations (driven by on-board data storage 
issues and downlink frequency), the \spitzer{} deep large area surveys must 
consist of multiple epochs of data with multiple pointings. For asteroid 
detections this is advantageous as the observation duration limit for all 
IRAC observations is 6~hrs which allows observations to be repeated with 
a cadence of a few hours. Assuming circular Keplerian orbits within the 
asteroid belt, the angular velocity $\omega$ (rad~s$^{-1}$), or rate of 
motion on the sky of an object near opposition is 

\begin{equation}
\omega=\sqrt{\frac{GM_{\odot}}{R^3}} 
\label{omega-eqn}
\end{equation}

\noindent where $G = 6.67 \times 10^{-11}~\rm{m}^{3}~\rm{kg}^{-1}~\rm{s}^{-2}$,
$M_{\odot}$ is the mass of the sun, and $R$ is the orbital distance in 
meters. For inner main belt asteroids with a semi-major axis of 2.2 AU, 
the expected rate of motion is 45\arcsec\ per hr, while the expected rate 
of motion for outer main belt asteroids is 24\arcsec \ per hr. However,
the actual rates and direction of motion on the sky may significantly
differ from these values (e.g., near quadrature, when the FLS fields
were observed at solar elongation $\simeq$ 115\degr). Nevertheless, movement
on the sky of many tens of arcsec per hr enables one to detect asteroids 
in the field on timescales of hours. 


The FLS consisted of IRAC observations of two 0.13 square degree  
fields (10\arcmin $\times$ 48\arcmin) centered on a solar elongation of 
115\degr\ as seen from \spitzer, at ecliptic latitudes $\beta = 
0$\degr\ and $+5$\degr\ on 2004 January 21~UT. In J2000 coordinates, this 
corresponds to field centers of R.A. $= 12^{h}03^{m}22.03^{s}$, decl. 
$= -00$\degr 21\arcmin 53.8\arcsec \ for the $\beta = 0\degr$ field, and RA 
$=12^{h} 11^{m} 20.10^{s}$, decl. $= +04$\degr 13\arcmin 18.7\arcsec \ for 
the $\beta = +5$\degr\ field as viewed by \spitzer. To detect asteroid 
motion, each IRAC field was observed at three epochs separated by 70 min. 

For each of the latitude fields, we colored each epoch and coadded the 
IRAC 8.0~\micron \ mosaics of the three epochs using a fixed stellar WCS to 
create a composite red, green, blue (RGB) image, which was then examined. 
In this RGB image, fixed sources appear white, and moving targets were 
seen as red-green-blue quasi-linear sequences of sources. The typical SNR 
for the detected sources was $\gtsimeq 10$. Using the RGB technique with 
the IRAC 8.0~\micron \ data, we were able to identify 18 sources in the 
$\beta =0$\degr \ field, and 16 sources in the $\beta = +5$\degr \ field. 
The faintest of these sources have 8.0~\micron \ fluxes of $\simeq 80~\mu$Jy 
and were detected with a SNR $\sim 10$ in both fields. For
comparison, \citet{vmeadows04} identified 15 known  (six of which we
also recover) and 19 unknown asteroids their analysis of 
the FLS fields, reporting 8.0~\micron \ fluxes ranging from 5.6 to 0.09~mJy.


The SWIRE program was designed to explore large scale structure in the 
universe with the IRAC and MIPS cameras \citep{lonsdale04}. This survey 
consisted of 6 fields selected to overlap those observed at depth in by 
the \textit{Hubble Space Telescope} and the \textit{Chandra X-Ray 
Observatory}.  Two epochs of observations were obtained for each of the 
six fields to preen moving targets from final mosaics enabling high 
fidelity point source catalogs to be generated. For the purposes of our 
asteroid survey we used IRAC data obtained between 2004 July 23 and 2004 
July~28 in the SWIRE XMM-LSS field, located at $\beta = -17$\degr \ and 
encompassing an area of 9.1 square degrees. Due to the large area covered 
by this \spitzer{} survey and duration limitations of single observations, 
the IRAC data consists of 16 individual pointings of 0.44 square degrees 
in two epochs. The two epochs of each pointing are offset by 3.75 hrs, 
thus significant asteroid motion can be detected. 

Asteroid detection in the SWIRE data sets was also performed by coloring and 
coadding IRAC 8.0~\micron \ mosaics of each epoch to create a composite 
red-blue (RB) image. In this RB image, fixed sources appear purple, and 
moving targets were seen as red-blue quasi-linear sequences of sources. We 
identify a total of 46 sources in the field. The faintest sources have 
8.0~\micron \ fluxes of $\simeq 300~\mu$Jy and were detected with a 
SNR $\sim 20$ in both fields. Of the 46 sources detected, only 
14 asteroids have ground based detections and orbital elements. 
 

Fluxes in all four IRAC channels were obtained for all asteroids in the 
FLS and SWIRE datasets. Pixel coordinates for each asteroid were recorded 
at the time of detection and were used for photometry. Though the IRAC 
cameras have an offset between the 3.6/5.8~\micron \ and the 
4.5/8.0~\micron \ FOVs, pipeline processed mosaics in each channel are 
aligned such that a given sky coordinate corresponds to the same pixel 
coordinate in all four channels. This allows use of one set of pixel 
coordinates for all sources in all 4 channels. 
Source photometry was performed using the IRAF task \textit{phot} and a 
three-pixel radius circular aperture to measure the asteroid flux and a 
concentric four-pixel wide sky annulus to determine the median sky 
contribution to the total flux measured in the integrated object aperture. 
Final fluxes and flux uncertainties were derived using the aperture 
corrections contained in the IRAC Data Handbook \citep{idh06}.
Due to the offset in the IRAC channel FOVs, some asteroids are only 
detected in the 4.6 and 8.0~\micron \ data. The number of asteroids 
detected in all four IRAC bands is 40 in the SWIRE data and 24 in the FLS 
fields. The mean (median) 8.0~\micron \ asteroid fluxes are 
$315 \pm 78~\mu$Jy ($199 \pm 78~\mu$Jy), $1276 \pm 680~\mu$Jy
($302 \pm 608~\mu$Jy), and $2448 \pm 53~\mu$Jy ($1355 \pm 53~\mu$Jy) for 
\spitzer{} FLS $0\degr$, FLS $5\degr$, and SWIRE fields respectively. The 
total number of asteroids detected in the \spitzer{} fields are 
summarized in Table~\ref{table:as_numsum}. 


Detailed physical parameters for the newly discovered asteroids 
detected at optical (58) and IR (41) wavelengths are presented in on-line
machine-readable catalogs. Illustrative catalog layouts are shown
in Table~\ref{table:machine-optical} and  Table~\ref{table:machine-spitz}.
The optical catalog, Table~\ref{table:machine-optical}, is similar to 
the standard Minor Planet Center submission 
format\footnote{\url{http://www.cfa.harvard.edu/iau/info/OpticalObs.html}}. 
In our Table~\ref{table:machine-optical} the corresponding column 
format is used: [1] the asteroid name, [2] a true/false=1/0 flag indicating 
if asteroid is a new discovery, [3]--[5] are respectively the year, 
month and UT date of observation, [6]--[11] are the right ascension 
and declination (J2000.0) of the detected target, [12] and [13] are 
the observed asteroid magnitude and associated uncertainty, while
[14] is the filter used for the observations. The IR 
machine-readable catalog, Table~\ref{table:machine-spitz}, is 
similar in column format where: [1] is the asteroid 
name, [2] is the \textit{Spitzer} Program ID number, [3] is the 
astronomical observation request (AOR) key, [4]--[9] is the year,
UT date and time at the start of an observation, [5] and [6] are the 
 right ascension and declination in degrees, [10]--[13] are
the right ascension and declination and  ecliptic latitude and longitude
in sexigesimal degrees, while [14] through [21] are the observed fluxes and
flux errors in $\mu$Jy measured for each asteroid in the four IRAC channels.
Note that the fluxes in the table are not color-corrected.

\section{DISCUSSION\label{disc}}


The distribution of asteroids within the main belt provides a means to
ascertain the mass distribution of material within the main belt and 
provides constraints for theoretical models developed to describe
planet formation from protoplanetary disks encompassing our
own proto-sun, as well as those extant around other young stellar systems.
By probing the asteroid distribution in reflected
optical light and in the thermal IR in coordinated 
observational programs, a more complete picture of the
number of asteroids and sizes and albedos can be obtained.  Current
models of solar system dynamics \citep[e.g.,][]{botk05} use the
distribution of absolute magnitudes of asteroids combined with a
single geometric albedo to estimate the size-frequency distribution of
asteroids as observationally determined diameters and albedos of main
belt asteroids only exist for $\sim$ 2400 objects. However, new survey 
data now available from fields observed with the \textit{Spitzer} 
can be used to determine asteroid albedos and diameters for sizes under 1~km,
enabling a more robust determination of albedo trends and size as a
function of latitude for a more representative population of objects
(e.g., a broader sampling of the total distribution).  Our study 
describes the first results for main belt asteroids with diameters of 
0.5~km.

\subsection{Completeness Limits}

To test completeness of our optical and \spitzer{} data sets, we used the 
IRAF task \textit{mkobjects} to create synthetic point sources in data 
frames. To assess completeness in the optical data, synthetic point 
sources were added with the same seeing values measured in each frame and 
with coordinates that varied to mimic asteroid motion. To model the 
\spitzer{} images, synthetic objects were randomly placed into an image 
with motions equivalent to the median rates of a body in Keplerian orbit 
at the corresponding ecliptic latitude. Synthetic asteroids were then 
recovered using the same RGB technique originally used to detected 
asteroids in the observational images. Our completeness values are not 
only sensitive to source fluxes, but also to motion. For valid 
completeness detection, an asteroid must be detected in at least 2 of the 
3 epochs of data. 

In our optical data, the 85$\%$ moving target completeness limit 
in all fields is $V = 21.3$.  Because the moving target completeness 
samples the completeness due to (1) a limiting flux and (2) motion, a 
point source completeness test was also performed. This exercise shows 
that the point source completeness at $V = 21.3$ is 90$\%$. For all 
asteroids counts at magnitudes greater than 21.3, the difference in the 
two completeness values of 5$\%$, is applied by assuming that the number 
of asteroids not detected due to motion is the same at all magnitudes. 
These completeness values have been used to obtain corrected asteroid 
number counts which are reported in column [3] in 
Table~\ref{table:as_numsum}. Figure~\ref{fig:lbt_histdist_4panel_fig1a}
shows the number counts per square degree as a function of magnitude in 
each optical field. 

The \spitzer{} IRAC~8.0~\micron \ analysis, under the constraints that the 
synthetic asteroid rate of motion was comparable to the median rates of 
motion in a field at a given ecliptic latitude combined with the 
requirement of a minimum of 2 detections per synthetic asteroid yields an 
80$\%$ completeness limit for the FLS and SWIRE data for $F_{8\micron} = 
103$~$\mu$Jy.  Our derived completeness value differs somewhat from that 
reported for the FLS by \citet{vmeadows04} who cite 90$\%$ completeness 
at 100~$\mu$Jy based on fixed target detections. With the completeness 
limit at 103~$\mu$Jy, we calculate that the number of asteroids in the 
FLS $0\degr$ and $5\degr$ fields are $148 \pm 12$ and $133 \pm 12$ per 
square degree respectively, and $5\pm 2$ per square degree for the 
$-17\degr$ SWIRE field. Figure~\ref{fig:spitzer_histdist_3panel_fig2} 
are histogram plots of the surface density of asteroids as a function of 
8.0~\micron \ flux in the \spitzer{} FLS and SWIRE fields. 

\subsection{Bulk size-frequency distribution\label{bsz-frq}}


Comparison of these two major data sets in different wavelengths is
complicated in part by the potential for a disparate probe of asteroid
sizes. Using the completeness limits in the $V$--band and at
IRAC~8.0~\micron\ we can estimate the minimum radii of asteroids detectable
in our survey from

\begin{equation}
F_{reflected}=\frac{2h \nu^{3}}{c^{2}}\frac{1}{e^{\frac{h \nu}{kT_{\odot}}}-1}
\frac{R_{\odot}}{r} \frac{R p_{v} \pi {D}^{2} cos(\tilde{\alpha})}
{16 \pi {\Delta}^{2}}
\end{equation}

\noindent and

\begin{equation}
F_{thermal}=\frac{2 \epsilon D^{2} h \nu^{3}}{{d_{Spitzer}}^{2} c^{2}}
\left[ \int_0^{\frac{\pi}{2}} \int_{\frac{-\pi}{2}}^{\frac{\pi}{2}}
\frac{1}{e^{\frac{h\nu}{k T_{ast}}}-1} cos^{2}(\phi)
cos(\theta-\alpha)\, d\theta \, d\phi \right]
\label{equation:eq-fthermal}
\end{equation}

\noindent where
\begin{equation}
T_{ast}= \left[\frac{(1-A)S_{\odot}}{r_{h} \eta \epsilon 
\sigma_{SB}} \right]^{\frac{1}{4}} (cos \phi)^{\frac{1}{4}}\,
(cos \theta)^{\frac{1}{4}}.
\end{equation}

\noindent The terms $F_{reflected}$ and $F_{thermal}$ are the 
observed $V$--band and IRAC 8.0~\micron \ band fluxes (W~m$^{-2}$~Hz) 
respectively. Variables in the
equation for reflected flux are: $\nu$ is the frequency of
observation, in this case $V$--band, $R_{\odot}$ and $T_{\odot}$ are the
radius (m) and temperature (K) of the Sun, $r$ is the asteroid
heliocentric distance in meters, $D$ is the asteroid diameter in meters
and $\Delta$ is the asteroid geocentric distance, $p_{v}$ is the
geometric albedo of the asteroid in $V$--band, R is the relative
reflectance of the asteroid as measured with respect to the $V$--band
reflectance and $\tilde{\alpha}$ is the observed phase angle in
radians. Other variables in the equation for thermal flux are:
$d_{Spitzer}$ is the asteroid-\spitzer{} separation (AU), $\epsilon$
is the asteroid emissivity which is assumed to be 0.9, A is the Bond
Albedo, $S_{\odot}$ is the solar constant, $r_{h}$ is the asteroid
heliocentric distance in AU, and $\eta$ is the beaming parameter.
Adopting mean values for the known asteroids in our datasets
$r_{h} = 2.82$~AU, $\delta = 1.89$~AU (optical) and
$r_{h} = 2.68$~AU, $d_{Spitzer} = 2.28$~AU, (\textit{Spitzer}
asteroids; Table~\ref{table:stmres}), assuming $\eta = 1$ with $p_{v} = 
0.15$ and $cos(\tilde{\alpha}) = 0.99$, the 
minimum asteroid radii detected in our data are $\simeq 1020$~m 
in the optical and $\simeq 450$~m in the IRAC 8.0~\micron \ channel. 
Although \spitzer{} is far more efficient than optical
imaging surveys at detecting smaller asteroids, asteroids identified from
these two data sets comprise a set of bodies with similar size ranges
within the main belt.


The IR and optical data can be used to obtain the power law 
size-frequency distribution at 3 similar latitudes. In both the optical and 
the mid-IR, flux can be used as a proxy for size in obtaining the 
size-frequency distribution assuming that the Bond albedo is well behaved 
as a function of wavelength. By adopting a power law distribution for the 
surface density distribution $\propto$ $f^{- \alpha}$, we use the 
completeness corrected counts to measure the power law slopes summarized 
in Table~\ref{table:oir-slopes}.  The derived optical and IR slopes are 
similar at all latitudes within the uncertainties, with a mean
value of $\alpha = -0.111 \pm 0.077$ for the surface density distribution
of asteroids. However, the formal errors suggest that the optical data
provides a better constraint on $\alpha$. Our derived $\alpha$-values 
(Table~\ref{table:oir-slopes}) for the \spitzer{} fields differs somewhat 
from those cited by \citet{vmeadows04}. The differences can be attributed to 
how the asteroid photometry was conducted; \citet{vmeadows04} employed 
PSF-fitting (using, at that time, a poorly determined IRAC PSF), whereas we 
resort to aperture photometry to measure the flux density. In addition, 
differences in completeness also lead to variances in our results as 
opposed to those presented in \citet{vmeadows04}. We cite 80\% completeness at 
100~$\mu$Jys for moving-objects, while \citet{vmeadows04} quote 
90\% completeness at the same flux level for point sources.

As the derived slopes of the surface density distributions are similar at 
optical wavelengths, we employed a Kolmogorov-Smirnov (KS) test using the 
mean rates of motion of the asteroids at $0\degr$ and $15\degr$ ecliptic 
latitude to discriminate whether these bodies detected in the  optical 
fields are drawn from the same size and magnitude population 
distributions. Unlike a $\chi^{2}$ test, data is not binned in the KS 
test and the KS test can be used on data sets of different lengths. Due 
to the low number of asteroids detected in the 10$\degr$ and 15$\degr$, 
binning of the data was rejected due to the associated loss of 
information and arbitrariness of defining bin sizes. The derived 
probability is 96.88\%, strongly suggesting that the two fields represent 
the same distance (rate) population. Similarly, a KS test to determine 
the likelihood that the asteroids represent the same size distribution 
yields only a 41.32\% probability that the objects at $15\degr$ represent 
the same size distribution as those detected at $0\degr$ ecliptic 
latitude. This statistical inference suggests that the size distribution 
of asteroids at $0\degr$ and $15\degr$ are dissimilar; however, the 
veracity of this conclusion is limited by small number detection 
statistics of our data. Table~\ref{table:ks-mmm} summarizes our 
probability analysis of the magnitude distribution (e.g., asteroid 
diameters) as a function of ecliptic latitude compared to the $0\degr$ 
field. Lastly, although the KS test indicates that the $0\degr$ 
and $15\degr$ optical fields are likely not the same population, they do 
appear to follow a distribution in which large, thus bright, objects are 
rare, whereas small and thus faint, objects are more numerous. 

\subsection{Comparison to models\label{sf-comp-models}}

Our observed asteroid frequency distribution can be juxtaposed 
with current models describing the evolution of the asteroid populations 
in the solar system, such as the Statistical Asteroid Model (SAM) 
developed by \cite{tcz05}. The SAM uses a set of 8603 asteroids with 
absolute magnitudes less than 15.75 (asteroids with diameters of 
$\gtsimeq$ 2.5 km) to determine a size-frequency distribution of 
asteroids which is assumed to be smooth to diameters of 1 km. This 
model also incorporates an albedo distribution derived from 15 dynamical 
families and 3 ``background'' populations based on 1980 asteroids 
which have diameter and albedo determinations from MSX \citep{msx02} and 
IRAS \citep{vt92, iras02}. A number of asteroids are excluded from this model, 
including those with inclinations greater than $25\degr$, those with 
eccentricities $>$ 0.3. 


The asteroid number counts observed with both \spitzer{} and the LBT are
less than those predicted by the SAM.  For a limiting flux of 0.06~mJy at
8~\micron \ the SAM predicts $430 \pm 40$ asteroids at $0\degr$ latitude
and $250 \pm 20$ asteroids at $5\degr$ latitude. One reason the SAM may
overestimate the number of asteroids is the assumed faint flux limit of
\citet{tcz05} which corresponds to an effective diameter of 0.6~km.
This flux limit is fainter than the completeness for
both fields, thus the SAM over predicts the number of small asteroids which
would be observed by \spitzer. The overestimate of the number of
asteroids may also be due to a fundamental assumption of the SAM --  the
power law slope of the size-frequency distribution is continuous
to diameters less than 1~km. This is not observed in the optical. The
size-frequency distribution appears to fall off around 3~km \citep{jm98};
however, the reasons for this fall off are not well understood. Two
possible reasons for the fall off in observed data are: 1)~the
bright limiting magnitude of most asteroid surveys excludes detections of
asteroids smaller than 1~km in diameter, or 2)~the asteroids smaller than
1~km may not be solid, but may instead be easily disruptible piles of
rubble. If the predicted number of asteroids is scaled from the
SAM with a model that yields the number of
asteroids with diameters $>$ 0.6~km being 2.4 times the number of
asteroids with diameters $>$ 1.0~km, the SAM still predicts $179 \pm
17$ asteroids in the $0\degr$ latitude field and $104 \pm 8$ asteroids
in the $5\degr$ latitude field. The estimate at the high latitude is
nearly in agreement with our FLS results; however, the SAM still
overestimates the number of asteroids in the $0\degr$ field. The latter
discrepancy may due to asteroids with high albedos ($>$ 0.25) which are
not detected in the IRAC channels because they are too cool and thus
their thermal fluxes are below the detection thresholds in this data.


For the LBT data which has limiting diameters of $\sim$ 2~km, the number 
counts are within the uncertainty of the SAM prediction. For the 
SDSS survey which has a point source completeness limit of $V = 22.2$, the 
SAM predicts $115 \pm 10$ asteroids at $0\degr$ latitude and $75 \pm 5$ 
asteroids at $5\degr$ latitude.  This completeness limit is 
fainter than ours by 0.9~mag; however, the model values are 
nearly in agreement with the asteroid number counts at these two 
latitudes. Unfortunately, the SAM does not predict asteroid 
counts at higher latitudes ($\beta > 5$\degr) and the
current version of the SAM only uses asteroids with inclinations less 
than $20\degr$ to make estimates on number counts. A subset of
our newly detected asteroids have large inclinations. In fact this
limitation of the SAM complicates derivation of the
size-frequency distribution and physical characteristics of
well known and well characterized targets such as 
Pallas and asteroid dynamical groups such as Hungarias and 
Phocaeas \citep{carv01}. 

\subsection{Asteroid Diameters, Albedos, and Colors\label{stm-da}}

Direct comparison of mean and median IRAC 8.0~\micron \ fluxes
(\S\ref{saf}) of asteroids in the two FLS fields and the SWIRE field,
reveals that the SWIRE asteroids are, on average across the population,
brighter than those asteroids at $0\degr$ and $5\degr$ latitude detected
in the FLS fields, by a factor of two. This variance in brightness could
be explained as a difference in asteroid diameter assuming that the
populations have comparable albedos. We explored this possibility by
estimating the diameters and deriving albedos for all asteroids which
were detected in all four IRAC bands using the Near Earth Asteroid
Thermal Model \citep[NEATM;][]{del02, delbo04}. In this model, the
thermal flux is dependent upon the sub-solar temperature and the
temperature distribution of the surface of the asteroid. The latter
is solely dependent on the albedo and $\eta$ as model variables. 
Thus, our implementation of NEATM follows the \citet{del02}
and \citet{delbo04} prescription whereby the model
fits the IR fluxes and the optical absolute magnitude of asteroids 
($H$) by varying the geometric albedo (through a $\chi^{2}$-minimization)
until a best fit is found. The \spitzer{} fluxes
were first color-corrected according to the prescription described in
the IRAC Data Handbook \citep{idh06}, resulting in the measured
fluxes being divided by 1.1717 and 1.1215 at 5.8~\micron \ and
8.0~\micron, respectively. The asteroid diameter is then calculated
by the relation of \citet{fc92} which uses only the geometric albedo
and the absolute magnitude as input values. Because IRAC asteroid spectral
energy distributions are composites of both thermal emission and 
reflected solar light \citep[e.g.,][]{mueller07}, we derived albedo and 
diameter estimates of the asteroids using only data from IRAC channels 3
and 4.  From the channel 3 and 4 fits, we find that the percentage of 
reflected solar light in our channel 1 photometry to be 
$\simeq$ 56\% and $\simeq$ 14\% in the channel 2 photometry.  Our 
modeling results for {\it known} asteroids detected in our IRAC fields 
are summarized in Table~\ref{table:stmres}. The mean diameters
of asteroids in SWIRE is 5.91~km while the mean asteroid diameter
in the FLS is 2.28~km.

Analysis of the mean geometric albedos and derived sizes for these 
objects suggests they divide between the two populations. The mean 
geometric albedo for asteroids in the FLS fields is 
$0.24\pm 0.07$ as opposed to $0.10\pm0.05$ for asteroids
in the SWIRE fields. No low albedo asteroids ($p_{V}$ $\ltsimeq$ 0.1) 
were detected in the \textit{Spitzer} FLS fields, whereas in the SWIRE 
fields they are frequent. Evidently the surface composition
of the asteroids in the SWIRE fields contains more carbonaceous
material than the asteroids in the FLS, as carbonaceous chondrites have
significantly lower albedos in the 0.03 to 0.11 range \citep{gaff76,jf73}.

However, the marked dichotomy in derived albedos is most likely
a result of the two \textit{Spitzer} surveys sampling different asteroid 
taxonomy classes within the general asteroid belt. An extensive 
analysis of $\sim 88,0000$ objects in the Sloan Digital Sky Survey 
Moving Object Catalog (SDSS MOC) 4 by \citet{parker08} demonstrates that 
there is a strong correlation between SDSS color, asteroid taxonomy 
(C-, S-, V-type), and orbital elements. In Fig.~\ref{fig:sdssmoc4_fig3}
we plot the proper orbital elements $a$ (orbital semi-major axis) vs.
$i$ (orbital inclination) of our SWIRE (green squares) and FLS (red triangles) 
detected asteroids overlaid on 207,942 numbered asteroids (black dots) with 
orbital elements in the ASTORB files \citep{bowell01}. The location of 
the major Kirkwood gaps as defined by \citet{parker08} are indicated, and 
the colored symbols used to identify the \textit{Spitzer}-detected 
asteroids are coded by derived albedo. Open symbols are for 
\textit{Spitzer} asteroids with derived albedos, $p_{V} \ltsimeq 0.1$, 
while filled symbols identify asteroids with $p_{V} > 0.1$. The majority 
($\simeq 80$\%) of the low albedo asteroids detected in the 
\textit{Spitzer} data reside at high inclination in the
outer belt. Primitive asteroids, with C-type colors and generally low
albedos ($p_{V} < 0.1$), are seen to dominate the asteroid population
in the outer belt, while S-type asteroids with $p_{V}$ ranging from 
$\sim 0.15$ to 0.2 tend to be more dominant in the middle and inner
asteroid belt \citep{parker08}. Our SWIRE data clearly samples asteroids in
the middle and outer belts, and in fact, for the first time,  we are able to 
estimate the albedos of km-sized asteroids in the outer belt
(Table~\ref{table:stmres}).

The low albedos ($p_{V} \ltsimeq 0.1$) of the $\simeq 6$~km-sized
outer belt SWIRE asteroids are similar to the albedos derived for
Jupiter-family comets, $0.02~\ltsimeq~p_{V}~\ltsimeq~0.06$
\citep{fernandez08, lamy04}. Recent modeling by \citet{levison08}
investigating the capture likelihood of cometary planetesimals into
the asteroid belt (leading the establishment of a D-type population)
suggests that the inner edge of the D-type population is near
$a \sim 2.6$~AU. Whether the population of low albedo asteroids discovered
in our SWIRE fields discussed here are organic-rich, 
primitive objects can only be verified with follow-up spectroscopy.



From flux measurements in all 4 IRAC channels, we can also begin to 
constrain the colors of asteroids. This is beneficial for both the 
understanding of bulk asteroid albedo variations as a function of size 
and other orbital parameters such as orbital semi-major axis or 
inclination and for the future exclusion of asteroids from galactic or 
extragalactic catalogs. Figure~\ref{fig:color_fig4} is an adaptation of 
the work by \citet{lacy04} which shows the IRAC colors of fixed 
astronomical objects where we have now included asteroid color data derived
from the our FLS and SWIRE field photometry. Asteroids mainly reside in one
isolated locus in color space. The size of this locus is dependent 
upon both temperature and albedo. The shortest wavelength observations 
with IRAC at 3.6~\micron \ trace the fraction of reflected incident solar 
flux, thus traces albedo, while the 8.0~\micron \  to 4.5~\micron \ ratio 
traces the thermal emission and thus the albedo and distance of an asteroid. 

\section{CONCLUSION\label{concl}}

Our combined optical survey of $\simeq 0.96$ square degrees and
\textit{Spitzer} IRAC~8.0~\micron{} fields encompassing from 0.26 to
$\simeq 9.1$ square degrees of sky at five latitudes perpendicular to the 
ecliptic plane resulted in detection of 118 main belt asteroids,
of which 91 are newly identified objects either at optical or infrared
wavelengths. The optical and mid-IR asteroid counts and fluxes demonstrate 
that the slope of the size-frequency distribution is consistent with that 
of a similar size population at 0$\degr$ and 15$\degr$ ecliptic latitude. 
The derived power law slopes of the asteroid surface density distribution are
similar at within each $\sim 5$\degr{} ecliptic latitude bin with a mean
slope of $-0.111 \pm 0.077$.  The observed asteroid number 
densities at optical wavelengths are generally consistent with those 
derived from the Standard Asteroid Model within the ecliptic plane. However, 
we find an over density at $\beta \gtsimeq 5$\degr{} in our optical fields,
while the infrared number densities are under dense by factors of 2 to 3
at all ecliptic latitudes.
 
For the 28 known asteroids detected in all four IRAC channels, mean albedos 
range from $0.24 \pm 0.07$ to $0.10 \pm 0.05$, statistically suggesting 
that these are two different populations. Our SWIRE data clearly 
samples low albedo asteroids in the middle and outer belts, and in fact, for 
the first time, we are able to estimate the albedos of km-sized asteroids 
in the outer belt.

In order to increase our number statistics and lower the errors, we are
completing the data-mining of additional, extant  multi-epoch survey 
\spitzer{} IRAC and MIPS dataset and are undertaking complimentary
ground based optical observations with the LBT and the 2.3-m Bok 
Telescope at Kitt Peak. These data will provide a survey area more 
than 3 times larger than that described in this manuscript and will 
also expand analysis of asteroid number counts to ecliptic latitudes of 
20\degr.

The errors in the slope of the size-frequency distribution cannot be 
decreased by greater number statistics alone.  The use of fluxes as 
proxies for asteroid sizes does not account for variations in albedo; 
therefore, additional mid-IR data is required to obtain albedos 
and sizes for a large number of asteroids. We in the process of
completing the data-mining of additional, extant multi-epoch survey 
\spitzer{} IRAC and MIPS dataset while undertaking complimentary
ground based optical observations with the LBT and the 2.3-m Bok
Telescope at Kitt Peak in order to increase our number statistics 
and lower the completeness statistics uncertainties. These data will 
provide a ecliptic survey area more than 3 times larger than that 
described in this manuscript and will also expand analysis of 
asteroid number counts to ecliptic latitudes of 20\degr.
These data will extend the number of asteroids with accurate sizes to 
diameters much smaller than the 10~km IRAS detection limit enabling
reliable number counts for asteroids with diameters 
$\ltsimeq 1$~km and assessment of whether the fall off in the size-frequency 
distribution of main belt asteroids distribution with 
diameters $\ltsimeq 1$~km is a real signature, or an artifact 
introduced by the limiting magnitudes of current asteroids surveys. 
  
\acknowledgements

This work is based on observations made with the {\it Spitzer} Space 
Telescope, which is operated by the Jet Propulsion Laboratory, California 
Institute of Technology under a contract with NASA. Support for this work 
was provided by NASA through an award issued by JPL/Caltech. Support for 
this work also was provided by NASA through contracts 1263741, 1256406, 
and 1215746 issued by JPL/Caltech to the University of Minnesota.
E.R. and C.E.W. also acknowledge support from the National Science Foundation 
grant AST-0706980. This work is also based on data acquired using 
the Large Binocular Telescope (LBT). The LBT is 
an international collaboration among institutions in the United States, 
Italy and Germany. LBT Corporation partners are: The University of Arizona 
on behalf of the Arizona University system; Istituto Nazionale di 
Astrophysica, Italy; LBT Beteiligungsgesellschaft, Germany, representing the 
Max-Planck Society, the Astrophysical Institute Potsdam, and Heidelberg 
University; The Ohio State University, and The Research Corporation, on 
behalf of The University of Notre Dame, University of Minnesota, and 
University of Virginia. The authors also acknowledge the detailed and
very helpful comments of anonymous referees whose insights greatly 
improved the manuscript.

{\it Facilities:} \facility{Spitzer (IRAC, MIPS)}
                  \facility{LBTO (LBC 4048 CCD)}

\clearpage

\clearpage

%
%


\begin{deluxetable}{lcccccc}
\tablewidth{0pt}
\rotate
\tablecaption{LBT/LBC OBSERVATIONAL SUMMARY\label{table:lbtobs}}
\tablecolumns{7}
\tablehead{
&\colhead{Field}&\colhead{Field}&&&&\colhead{Zero-point}\\
\colhead{Pointing Name} 
&\colhead{Center}
&\colhead{Center} 
&\colhead{Date} 
&\colhead{Time} 
&\colhead{Seeing} 
&\colhead{Offset}\\
\colhead{} 
&\colhead{RA (J2000)} 
&\colhead{Dec (J2000)} 
&\colhead{(2007 UT)} 
&\colhead{(UT)} 
&\colhead{($\arcsec$)}
&\colhead{(mag)}
}
\startdata
\underbar{0 degree field, pointing 1}\\
Epoch 1 & 6:14:21.68 & +23:25:15.49 & Jan 16 & 07:55:20 & 2.13& 
7.37$\pm$0.10\\

Epoch 2 & 6:14:21.68 & +23:25:15.49 & Jan 16 & 08:00:01 & 2.74& 
7.50$\pm$0.11\\

Epoch 3 & 6:14:21.68 & +23:25:15.49 & Jan 16 & 09:51:30 & 2.52& 
7.40$\pm$0.13\\

\underbar{0 degree field, pointing 2}\\
Epoch 1 & 6:14:22.33 & +23:40:18.72 & Jan 16 & 08:01:21 & 2.19& 
7.68$\pm$0.11\\

Epoch 2 & 6:14:22.33 & +23:40:18.72 & Jan 16 & 09:13:54 & 2.67& 
7.81$\pm$0.06\\

Epoch 3 & 6:14:22.33 & +23:40:18.72 & Jan 16 & 09:57:22 & 2.67& 
7.78$\pm$0.06\\
\\

\underbar{5 degree field, pointing 1}\\
Epoch 1 & 6:14:56.16 & +28:25:17.94 & Jan 25 & 02:49:41 & 1.50& 
7.43$\pm$0.04\\

Epoch 2 & 6:14:56.16 & +28:25:17.94 & Jan 25 & 03:49:01 & 1.64& 
7.47$\pm$0.04\\

Epoch 3 & 6:14:56.16 & +28:25:17.94 & Jan 25 & 06:39:50 & 2.20& 
7.66$\pm$0.04\\

\underbar{5 degree field, pointing 2}\\
Epoch 1 & 6:14:56.78 & +28:40:19.50 & Jan 25 & 02:55:22 & 1.50& 
7.48$\pm$0.04\\

Epoch 2 & 6:14:56.78 & +28:40:19.50 & Jan 25 & 03:54:47 & 1.75& 
7.56$\pm$0.04\\

Epoch 3 & 6:14:56.78 & +28:40:19.50 & Jan 25 & 06:45:37 & 1.71& 
7.66$\pm$0.04\\
\\

\underbar{10 degree field, pointing1}\\
Epoch 1 & 6:15:33.49 & +33:25:15.14 & Jan 24 & 02:17:27 & 1.44& 
7.36$\pm$0.03\\

Epoch 2 & 6:15:33.49 & +33:25:15.14 & Jan 24 & 04:34:49 & 2.12& 
7.62$\pm$0.03\\

Epoch 3 & 6:15:33.49 & +33:25:15.14 & Jan 24 & 08:10:00 & 2.64& 
7.67$\pm$0.03\\

\underbar{10 degree field, pointing 2}\\
Epoch 1 & 6:15:34.08 & + 33:40:19.96 & Jan 24 & 02:23:01 & 1.44& 
7.33$\pm$0.03\\

Epoch 2 & 6:15:34.08 & + 33:40:19.96 & Jan 24 & 04:40:57 & 2.16& 
7.55$\pm$0.03\\

Epoch 3 & 6:15:34.08 & + 33:40:19.96 & Jan 24 & 08:15:42 & 2.65& 
7.55$\pm$0.03\\
\\

\underbar{15 degree field, pointing 1}\\
Epoch 1 & 6:16:17.02 & +38:25:09.47 & Jan 24 & 02:05:07 & 1.35& 
7.35$\pm$0.03\\

Epoch 2 & 6:16:17.02 & +38:25:09.47 & Jan 24 & 04:13:06 & 1.79& 
7.60$\pm$0.02\\

Epoch 3 & 6:16:17.02 & +38:25:09.47 & Jan 24 & 07:47:27 & 2.62& 
7.68$\pm$0.03\\

\underbar{15 degree field, pointing 2}\\
Epoch 1 & 6:16:15.22 & +38:40:11.62 & Jan 24 & 02:10:49 & 1.28& 
7.24$\pm$0.03\\

Epoch 2 & 6:16:15.22 & +38:40:11.62 & Jan 24 & 04:19:06 & 2.25& 
7.60$\pm$0.03\\

Epoch 3 & 6:16:15.22 & +38:40:11.62 & Jan 24 & 07:54:28 & 3.08& 
7.64$\pm$0.03\\

\enddata
\end{deluxetable}
\clearpage


\begin{deluxetable}{lcccc}
\tablewidth{0pt}
\tablecaption{SUMMARY OF ASTEROID DETECTIONS BY LATITUDE AND NUMBER COUNT 
STATITICS\label{table:as_numsum}} 
\tablecolumns{5}
\tablehead{
\colhead{Field} 
&\colhead{Number} 
&\colhead{Number} 
&\colhead{Total Number}
&\colhead{Limiting}\\ 
\colhead{Ecliptic} 
&\colhead{Asteroids} 
&\colhead{Asteroids}
&\colhead{Asteroids}
&\colhead{Magnitude}\\
\colhead{Latitude}
&\colhead{Detected}
&\colhead{Per Square Degree\tablenotemark{a}}
&\colhead{Per Square Degree\tablenotemark{b}}
&\colhead{or Flux}
}
\startdata
\cutinhead{LBT $V$--Band}
$0\degr$  & 24 & 85 & 91$\pm$\phantom{0}10 & 22.3\\
$5\degr$  & 21 & 74 & 80$\pm$\phantom{0}9 & 22.0\\
$10\degr$ & 11 & 39 & 41$\pm$\phantom{0}6 & 22.2\\
$15\degr$ & 6 & 21 & 23$\pm$\phantom{0}5 & 22.2\\

\cutinhead{FLS IRAC 8\micron}
$0\degr$  & 24 & 85 & 91$\pm$\phantom{0}10 & 25.7 $\mu$Jy\\
$5\degr$  & 21 & 74 & 80$\pm$\phantom{0}9 & 25.7 $\mu$Jy\\

\cutinhead{SWIRE IRAC 8\micron}
$-17\degr$& 11 & 39 & 41$\pm$\phantom{0}6 & 35.4 $\mu$Jy\\
\enddata
\tablenotetext{a}{Directly calculated from the number of asteroids detected
where each LBT field covers 0.24 square degrees of sky, the FLS
fields are 0.13 square degrees, and the single SWIRE field is 9.1 square
degrees in extent.}
\tablenotetext{b}{Number of asteroids including completeness.}
\end{deluxetable}
\clearpage


\begin{deluxetable}{lccccccccccccr}
\tablewidth{0pt}
\rotate
\tablecaption{OPTICALLY DETECTED ASTEROIDS -- MACHINE READABLE 
TABLE\tablenotemark{a}\label{table:machine-optical}}
\tablecolumns{14}
\tablehead{
\colhead{Asteroid}
&\colhead{Discovery}
&\colhead{Year}
&\colhead{Month}
&\colhead{Day}
&\colhead{RA\tablenotemark{a}}
&\colhead{RA}
&\colhead{RA}
&\colhead{Dec\tablenotemark{b}}
&\colhead{Dec}
&\colhead{Dec}
&\colhead{}
&\colhead{Mag}\\

\colhead{Name}
&\colhead{Flag}
&\colhead{}
&\colhead{}
&\colhead{(UT)}
&\colhead{(hrs)}
&\colhead{(min)}
&\colhead{(sec)}
&\colhead{(degr)}
&\colhead{(min)}
&\colhead{(sec)}
&\colhead{Mag}
&\colhead{Error}
&\colhead{Filter}
}
\startdata
sdt01&1&2007&01&16.33147&06&14&03.68&+23&21&03.0&19.72 &0.03&V\\
sdt01&1&2007&01&16.38195&06&14&01.21&+23&21&08.7&20.34&0.04&V\\
sdt01&1&2007&01&16.41215&06&13&59.70&+23&21&12.1&20.16&0.04&V\\
sdt02&1&2007&01&16.33147&06&13&56.23&+23&15&27.2&20.94&0.13&V\\
sdt02&1&2007&01&16.38195&06&13&53.82&+23&15&26.8&21.54&0.10&V\\
sdt02&1&2007&01&16.41215&06&13&52.35&+23&15&27.0&21.46&0.04&V\\
$\vdots$&\\
\enddata

\tablenotetext{a}{Full catalog available on-line.}
\tablenotetext{b}{RA and Declination given in J2000 coordinates.}
\end{deluxetable}

\clearpage




\begin{deluxetable}{lcccccccccccccccccccc}
\tabletypesize{\tiny}
\tablewidth{0pt}
\rotate
\tablecaption{SPITZER DETECTED ASTEROIDS -- MACHINE READABLE
TABLE\tablenotemark{a, b}\label{table:machine-spitz}}
\tablecolumns{21}
\tablehead{
\colhead{Asteroid}
&\colhead{}
&\colhead{AOR}
&\colhead{}
&\colhead{}
&\colhead{}
&\colhead{}
&\colhead{}
&\colhead{}
&\colhead{RA}
&\colhead{Dec}
&\colhead{Ecliptic}
&\colhead{Ecliptic}\\

\colhead{Name}
&\colhead{PID}
&\colhead{Key}
&\colhead{YR}
&\colhead{MN}
&\colhead{Day}
&\colhead{Hr}
&\colhead{Min}
&\colhead{Sec}
&\colhead{J2000}
&\colhead{J2000}
&\colhead{Long}
&\colhead{Lat}
&\colhead{F$_{3.6}$}
&\colhead{F$^{err}_{3.6}$}
&\colhead{F$_{4.5}$}
&\colhead{F$^{err}_{4.5}$}
&\colhead{F$_{5.8}$}
&\colhead{F$^{err}_{5.8}$}
&\colhead{F$_{8.0}$}
&\colhead{F$^{err}_{8.0}$}\\

\colhead{}
&\colhead{}
&\colhead{}
&\colhead{}
&\colhead{}
&\colhead{(UT)}
&\colhead{}
&\colhead{}
&\colhead{}
&\colhead{(\degr)}
&\colhead{(\degr)}
&\colhead{(\degr)}
&\colhead{(\degr)}
&\colhead{($\mu$Jy)}
&\colhead{($\mu$Jy)}
&\colhead{($\mu$Jy)}
&\colhead{($\mu$Jy)}
&\colhead{($\mu$Jy)}
&\colhead{($\mu$Jy)}
&\colhead{($\mu$Jy)}
&\colhead{($\mu$Jy)}\\
}
\startdata
2001QY160& 98& 6095104& 2004&01&21&02&35& 25.728&181.0084& -0.0412&
180.9416& 0.3633&89.4&8.1& 18.9&14.4& 95.7&45.7&939.9&179.4\\

0h                   &98&6095104&2004&01&21&02&35&25.728&181.0051&
-0.1183&180.9693&0.2913&2.3&2.2& 7.7& 6.9& 65.3&  35.5&
548.8&132.8\\

2004BS160&  98&  6095104& 2004& 01& 21& 02& 35& 25.728& 181.0004&
-0.1816& 180.9902&   0.2313&    3.7&   3.6&   1.7&   1.7 &  52.5&
52.2&   103.2&  49.2\\

2000AX136&  98&  6095104& 2004& 01& 21& 02& 35& 25.728& 180.9245&
-0.1815& 180.9204&   0.2012&    4.8&   4.6& 103.7&  47.2&  240.9&
79.9&  2526.7& 305.6\\

$\vdots$&\\

\enddata
\tablenotetext{a}{Full catalog available on-line.}
\tablenotetext{b}{Observed fluxes, no IRAC color-correction applied.}
\end{deluxetable}

\clearpage
%
%
%
%

\begin{deluxetable}{lcccr}
\tablewidth{0pt}
\rotate
\tablecaption{OBSERVED ASTEROID IRAC 
FLUXES\tablenotemark{a}\label{table:obflux}}
\tablecolumns{5}
\tablehead{
\colhead{Provisional}\\

\colhead{Designation}
&\colhead{F$_{3.6}$}
&\colhead{F$_{4.5}$}
&\colhead{F$_{5.8}$}
&\colhead{F$_{8.0}$}\\

\colhead{}
&\colhead{($\mu$Jy)}
&\colhead{($\mu$Jy)}
&\colhead{($\mu$Jy)}
&\colhead{($\mu$Jy)}
}
\startdata
\cutinhead{SWIRE Asteroids}
2004TH222&    18.5$\pm$ 14.1&      53.6$\pm$ 30.6&   241.0$\pm$
80.3&     1682.6$\pm$    246.3\\
2005YV181 &    6.0$\pm$ 5.6&       16.23$\pm$ 12.8&   128.4$\pm$
55.2&      933.8$\pm$    179.0\\
2005YV50    & 16.7$\pm$ 14.2&      15.7$\pm$ 12.4&      64.5$\pm$     36.0&
 570.6$\pm$    136.2\\
1999WA1      &45.1$\pm$ 27.1&     113.4$\pm$ 50.0&   938.5$\pm$
172.8&    7314.0$\pm$    532.0\\
2004TO8    & 109.2$\pm$ 48.7&     259.0$\pm$ 82.6&  1684.4$\pm$
237.5&   10301.9$\pm$    634.3\\
2006AP1      & 5.8$\pm$ 5.4&       32.8$\pm$ 21.7&   263.8$\pm$
84.7&     1677.9$\pm$    245.7\\
2005WX162   & 10.4$\pm$8.9&        38.0$\pm$ 24.1&   311.5$\pm$     93.5&
 2287.0$\pm$    290.30\\
1999VY171   & 75.7$\pm$ 38.5&     383.9$\pm$ 103.8& 2867.7$\pm$
314.2&   18500.5$\pm$    856.9\\
1999JZ7      & 4.0$\pm$ 4.9&       25.5$\pm$ 18.1&   196.8$\pm$
71.2&     1512.8$\pm$    232.3\\
2000VP1      &15.9$\pm$ 12.7&      25.0$\pm$ 17.8&   113.9$\pm$
51.2&      775.1$\pm$    161.3\\
2005YH117   &  6.7$\pm$ 6.2&       48.0$\pm$ 28.4&   470.0$\pm$
118.0&    3570.0$\pm$    366.2\\
2002FE15     &12.0$\pm$ 10.1&     47.9$\pm$ 28.4&   355.7$\pm$
100.8&     4308.9$\pm$    403.4\\
2005YX74     &83.1$\pm$ 42.7&    118.5$\pm$ 52.0&   538.9$\pm$
127.5&     4465.2$\pm$    411.6\\
2002CN96     &15.5$\pm$ 12.3&     65.0$\pm$ 35.3&   584.1$\pm$
133.3&     3106.3$\pm$    343.6\\

\cutinhead{FLS Asteroids}
2001QY160\tablenotemark{b} &    8.9$\pm$ 7.8&       31.8$\pm$ 22.6
&   101.3$\pm$ 47.5&      905.4$\pm$ 175.8\\
2000AX136\tablenotemark{b}   & 114.8$\pm$50.4&     217.6$\pm$ 74.2
&   1228.2$\pm$ 200.3&    8993.1$\pm$ 593.3\\
2004BH160   &  9.1$\pm$ 9.3&       11.4$\pm$ 10.9&      11.5$\pm$     11.7&
 73.6$\pm$ 39.6\\
2001QD49\tablenotemark{b}     & 7.1$\pm$ 7.4&       11.2$\pm$ 12.2
&      16.1$\pm$     16.0& 109.9$\pm$ 51.3\\
2002WL7\tablenotemark{b}      &18.2$\pm$ 13.9&      50.0$\pm$ 29.1
&   497.6$\pm$ 121.4&    3239.4$\pm$ 347.8\\
1999FZ24\tablenotemark{b}    & 41.7$\pm$ 25.9&      45.2$\pm$ 27.3
&   211.0$\pm$ 74.1&     2116.3$\pm$ 277.6\\
2001RP137\tablenotemark{b}  &   8.3$\pm$ 7.6&       14.3$\pm$ 14.2
&      34.9$\pm$     24.1& 273.5$\pm$ 89.2\\
2004EU91     & 9.3$\pm$ 9.2&        7.2$\pm$  6.8&      35.7$\pm$     26.5&
116.5$\pm$ 53.3\\
2004CA105  &  61.7$\pm$ 35.8&      54.6$\pm$ 33.7&      59.9$\pm$     34.8&
 487.4$\pm$ 124.5\\
2002RG106   & 12.2$\pm$ 10.4&       5.3$\pm$  5.3&      22.9$\pm$     17.6&
 228.3$\pm$ 80.5\\
2004BU99    &  6.3$\pm$ 5.8&        7.9$\pm$  7.4&      40.3$\pm$     25.9&
398.9$\pm$ 110.9\\
2004BS160  &   8.1$\pm$ 8.6&        3.1$\pm$  3.1&      21.4$\pm$     30.7&
101.9$\pm$  49.0\\
2001SB182   & 14.1$\pm$ 11.9&      11.8$\pm$ 11.0&      32.6$\pm$     22.6&
 184.0$\pm$ 70.1\\
2004FG8      & 6.1$\pm$ 7.2&        7.7$\pm$  7.0&      25.5$\pm$     19.2&
185.5$\pm$  70.7\\
\enddata

\tablenotetext{a}{Observed fluxes, no IRAC color-correction applied.}
\tablenotetext{b}{Asteroids also reported in \citet{vmeadows04}.}
\end{deluxetable}
\clearpage


%

\begin{deluxetable}{lcccccccr}

\rotate
\tablecaption{STANDARD THERMAL (NEATM) MODEL DERIVED ASTEROID
DIAMETERS AND ALBEDOS \label{table:stmres}}
\rotate
\tablewidth{0pt}
\tablecolumns{9}
\tablehead{
\colhead{Provisional}
&\colhead{Heliocentric\tablenotemark{a}}
&\colhead{Spitzer\tablenotemark{a}}
&\colhead{Phase}
&\colhead{Absolute\tablenotemark{b}}
&\colhead{NEATM}
&\colhead{Geometric}
&\colhead {Beaming}
&\colhead{Model}\\

\colhead{Designation}
&\colhead{Distance}
&\colhead{Distance}
&\colhead{Angle}
&\colhead{Magnitude}
&\colhead{Diameter}
&\colhead{Albedo}
&\colhead{parameter}
&\colhead{$\chi^{2}$}\\

\colhead{}
&\colhead{(AU)}
&\colhead{(AU)}
&\colhead{(\degr)}
&\colhead{(H)}
&\colhead{(km)}
&\colhead{}
&\colhead{}
&\colhead{}
}
\startdata
\cutinhead{SWIRE Asteroids}
2004 TH222  & 2.08 &1.79&29.4&15.4& 2.83& 0.14 &1.69& 0.0002\\
2005 YV181  &3.27&3.09&18.20&15.5& 3.91 & 0.07 & 0.79& 0.0157\\
2005 YV50    &2.38&2.14&25.36&16.0& 2.45 & 0.12 & 1.66& 0.0087\\
1999 WA1     &2.82&2.60&21.25&12.7& 9.16 & 0.18 & 1.01& 0.0225 \\
2004 TO8      &2.66& 2.44& 22.57&14.5& 9.02& 0.03 & 1.04&0.5826\\
2006 AP1      & 2.91&2.71&20.53&15.5& 3.59 & 0.09 & 0.77&0.0035\\
2005 WX162&2.83&        2.61&21.20&15.3& 5.10 & 0.05 & 1.03&0.0141\\
1999 VY171 & 2.56&2.34&23.48&14.0& 10.92 & 0.04 & 1.04&0.6644\\
1999 JZ7       &2.33&2.08&26.02&14.0& 3.39 & 0.39 &1.35& 0.0308\\
2000 VP1      &2.43&2.21&24.83&15.7& 1.99 & 0.23 & 1.03&0.0143 \\
2005 YH117 &2.91&       2.73&20.57&15.4& 6.73& 0.03 &1.01& 0.0161 \\
2002 FE15    &3.39&3.25&17.54&14.5& 12.41 & 0.02 & 1.05&0.3090 \\
2005 YX74    &2.85&2.64&20.99&14.8&5.35 &0.08 &1.05&0.0859 \\
2002 CN96   &2.95&2.74&20.29&14.8&5.88&0.06&0.87&0.9232\\

\cutinhead{FLS Asteroids}
2001 QY160 & 2.96& 2.39& 17.74& 15.0& 3.26 & 0.17 & 1.02&0.0241\\
2000 AX136 & 2.40& 1.80&22.02& 13.1 & 6.89&0.21&1.33& 0.0002\\
2004 BH160 & 2.42&1.82&21.86& 18.4& 0.51 & 0.30 &1.03& 0.0001\\
2001 QD49 & 2.77&       2.19&18.98& 17.5& 0.94 & 0.20 & 1.00& 0.0048\\
2002 WL7 & 2.69& 2.11& 19.55& 15.3& 4.48 & 0.07 & 1.05& 0.1406\\
1999 FZ24 & 3.10& 2.53& 16.93& 13.8& 5.81 & 0.16 & 1.03& 0.1115\\
2001 RP137 & 2.93& 2.36&17.90& 16.7 &  1.72 & 0.13 & 1.02& 0.0012\\
2004 EU91 & 2.40& 1.80&22.04& 18.1& 0.51 & 0.39 & 0.76& 0.1778\\
2004 CA105 & 2.80&2.22&18.77& 14.7& 2.27 & 0.44 & 1.03& 0.0062\\
2002 RG106 & 2.69& 2.10& 19.58& 16.1& 1.30 & 0.38 & 1.045&  0.108\\
2004 BU99 & 2.17&  1.54& 24.53&  17.4   & 1.84 & 0.06 & 2.50& 0.0008\\
2004BS160&  2.41& 1.81& 21.99& 18.6& 0.57 & 0.20 & 1.01& 0.0178\\
2001SB182& 3.02& 2.46& 17.36& 15.6 & 1.39 & 0.53& 0.75& 0.0636\\
2004FG8 &2.01& 1.37& 26.61& 18.5& 0.74& 0.13&1.90&0.0001\\

\enddata
\tablenotetext{a}{Heliocentric, $r_{h}$, and \spitzer-asteroid distances 
at epoch of observation.}
\tablenotetext{b}{Absolute Magnitudes are values given in the JPL
Horizons database, \url{http://ssd.jpl.nasa.gov/?horizons}.}
\end{deluxetable}
\clearpage



\begin{deluxetable}{lccc}
\tablewidth{0pt}
\tablecaption{POWER LAW SLOPES OF THE SURFACE DENSITY DISTRIBUTIONS
\label{table:oir-slopes}}
\tablecolumns{4}
\tablehead{
&& \colhead{Ecliptic}
& \colhead{Slope} \\
\colhead{Dataset}
& \colhead{Bandpass}
& \colhead{Latitude}
& \colhead{$\alpha$}
}
\startdata
LBT &  $V$ & $0\degr$ & $-0.118 \pm 0.070$\\

LBT &  $V$ & $5\degr$ & $-0.161 \pm 0.074$\\

LBT &  $V$ &$15\degr$ & $-0.131 \pm 0.093$\\

FLS &  IRAC 8\micron & $0\degr$ & $-0.083 \pm 0.145$\\

FLS &  IRAC 8\micron & $5\degr$ & $-0.100 \pm 0.240$\\

SWIRE &IRAC 8\micron & $-17\degr$ &$-0.075 \pm 0.340$\\
\enddata
\end{deluxetable}
\clearpage


\begin{deluxetable}{cccc}
\tablewidth{0pt}
\tablecaption{ASTEROID POPULATION PROBABILITIES, MEAN AND MEDIAN 
MAGNITUDES\label{table:ks-mmm}}
\tablecolumns{4}
\tablehead{
\colhead{Field Ecliptic}\\
\colhead{Latitude}
&\colhead{Probability\tablenotemark{a}} 
&\colhead{Median} 
&\colhead{Mean}\\
\colhead{(\degr)}
&\colhead{(\%)}
&\colhead{(mag)}
&\colhead{(mag)}
}
\startdata
\phantom{0}0 & 100 & 20.6 & 20.45\\
\phantom{0}5 & 66.24 & 20.5 & 20.41\\
10 & 38.89 & 19.8 & 20.00\\
15 & 41.32 & 20.8 & 20.65\\
\enddata
\end{deluxetable}
\clearpage



\begin{figure}
\begin{center}
\includegraphics[angle=0,scale=.75]{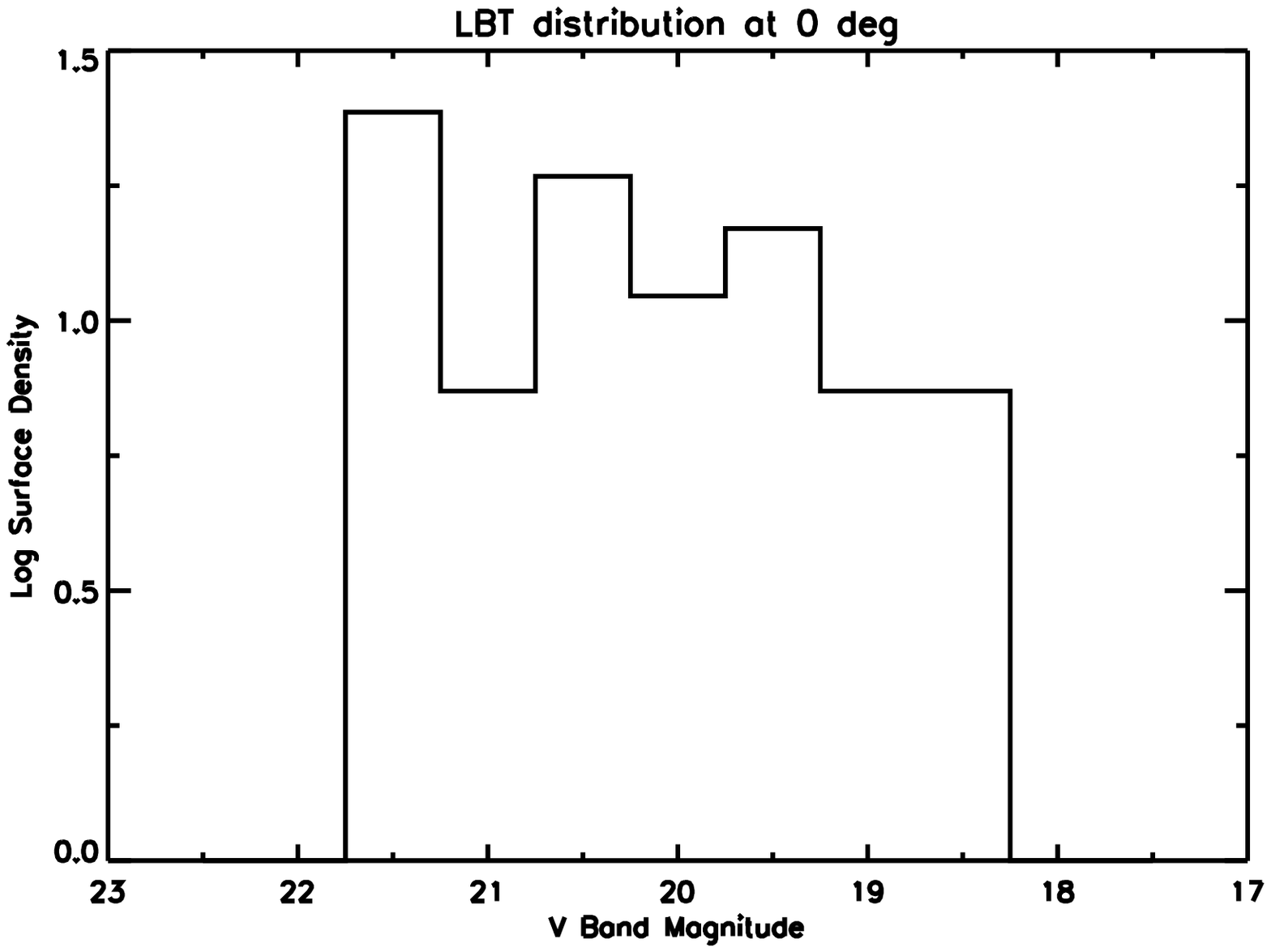}
\includegraphics[angle=0,scale=.75]{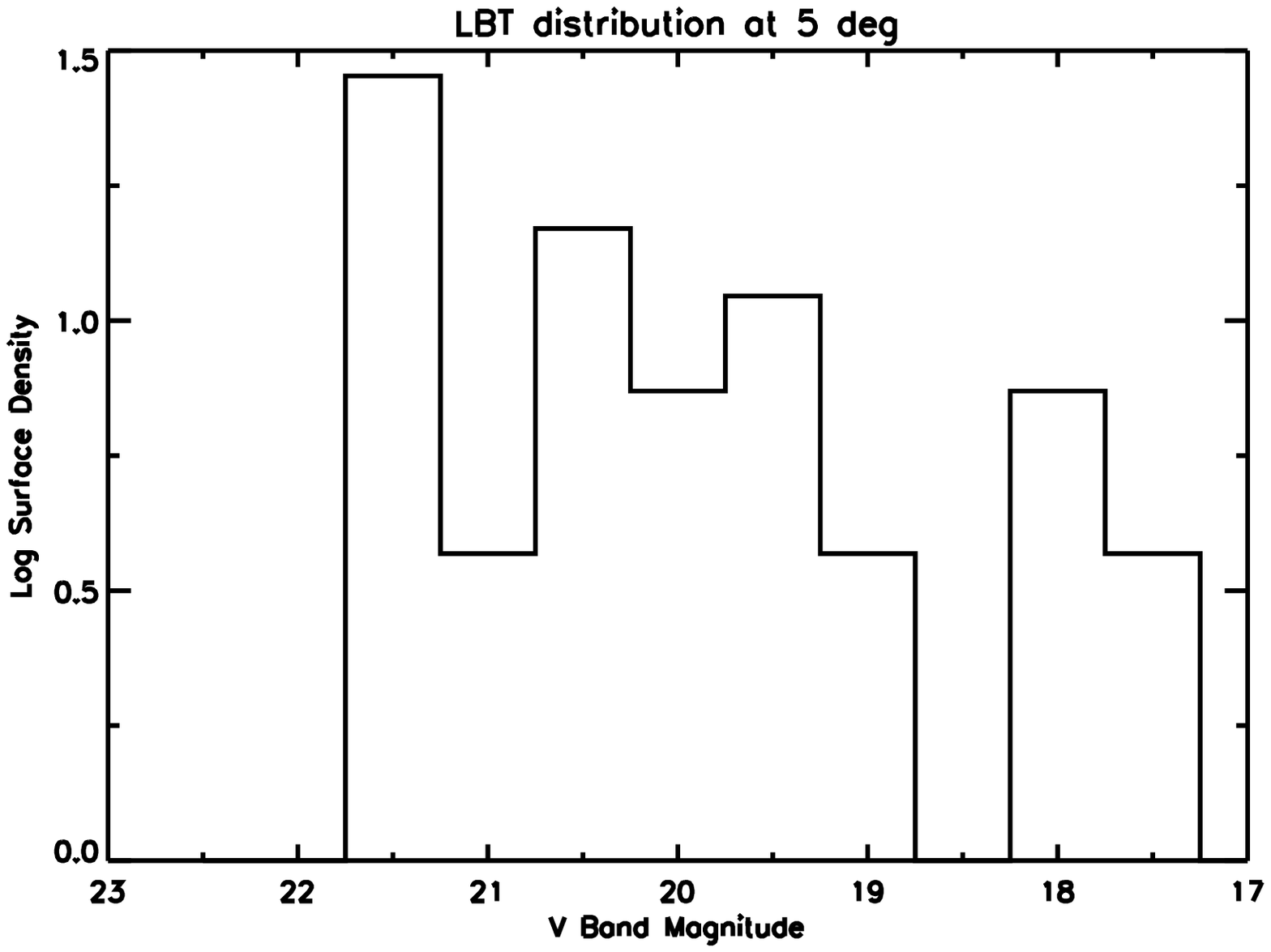}
\end{center}
\vspace{-0.75cm}
\caption{Magnitude histogram distribution for LBT asteroids at 
$0\degr$ ecliptic latitude ([a], upper) and $5\degr$ ecliptic 
latitude ([b], lower).
\label{fig:lbt_histdist_4panel_fig1a}}
\end{figure}
\clearpage

\setcounter{figure}{0}
\begin{figure}
\begin{center}
\includegraphics[angle=0,scale=.75]{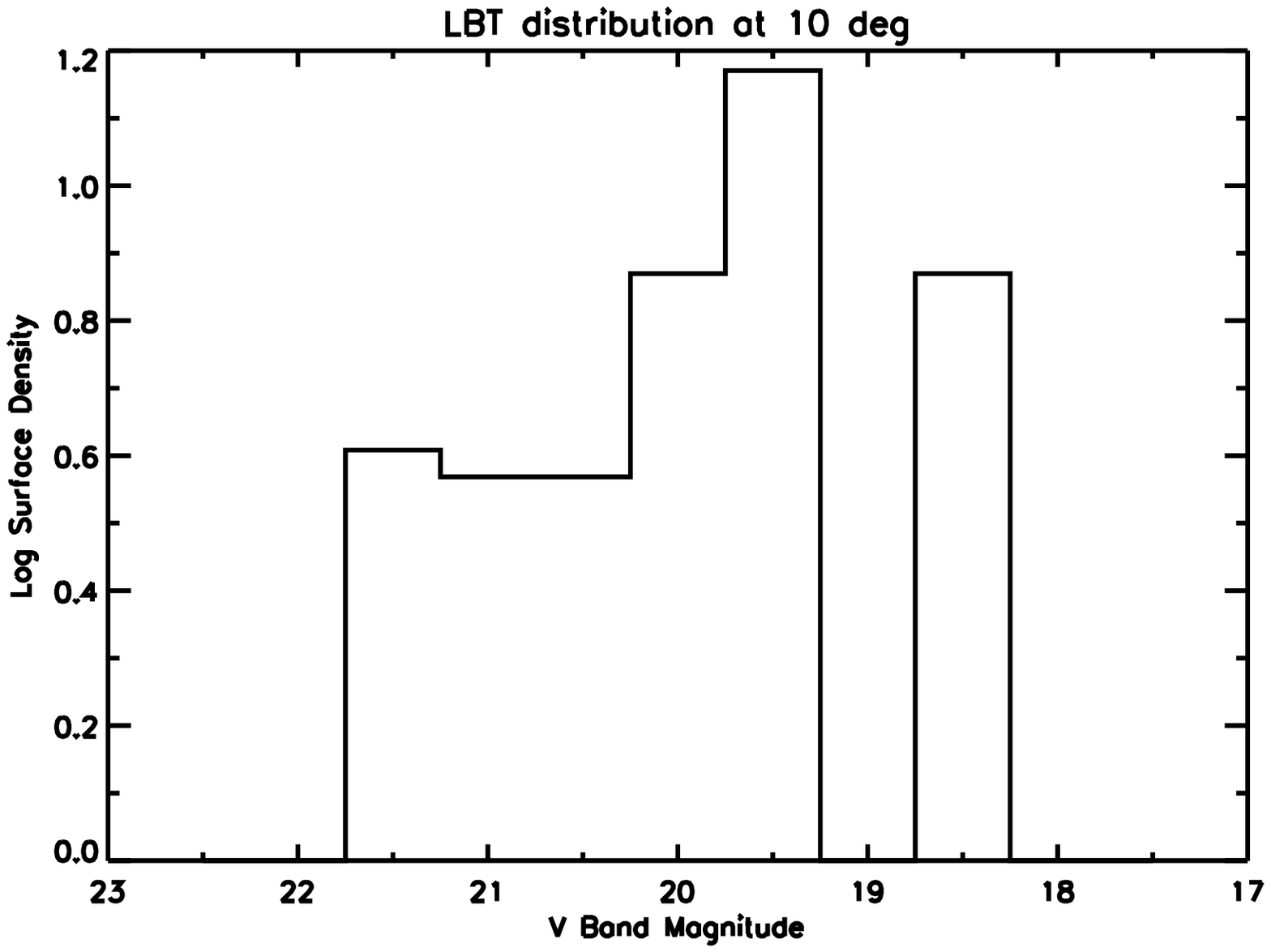}
\includegraphics[angle=0,scale=.75]{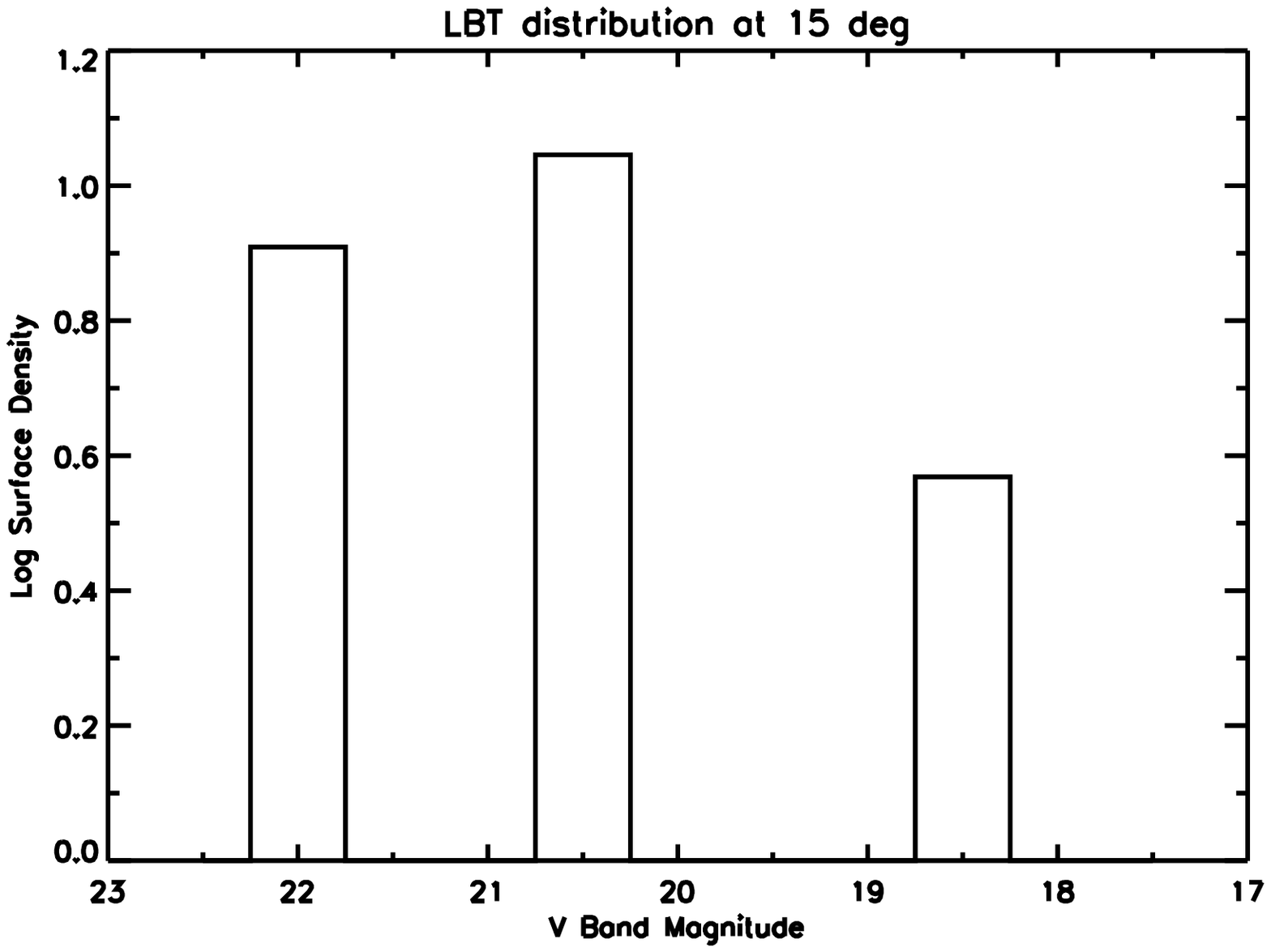}
\end{center}
\vspace{-0.75cm}
\caption{{\it continued:} Magnitude 
histogram distribution for LBT asteroids at $10\degr$ ecliptic latitude 
([c], upper) and $15\degr$ ecliptic latitude ([d], lower).
\label{fig:lbt_histdist_4panel_fig1b}}
\end{figure}
\clearpage


\begin{figure}
\begin{center}
\includegraphics[angle=0,scale=.55]{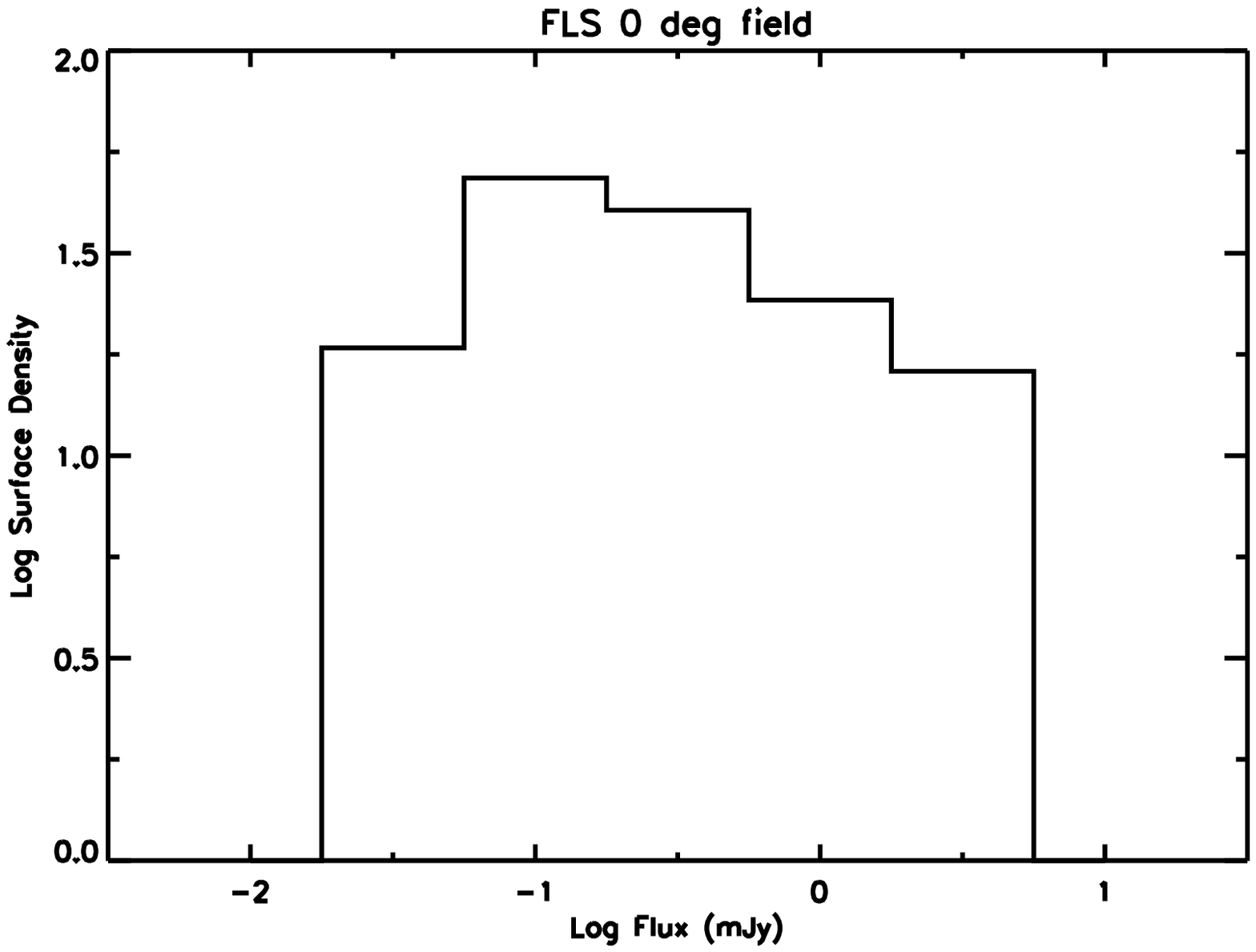}
\includegraphics[angle=0,scale=.55]{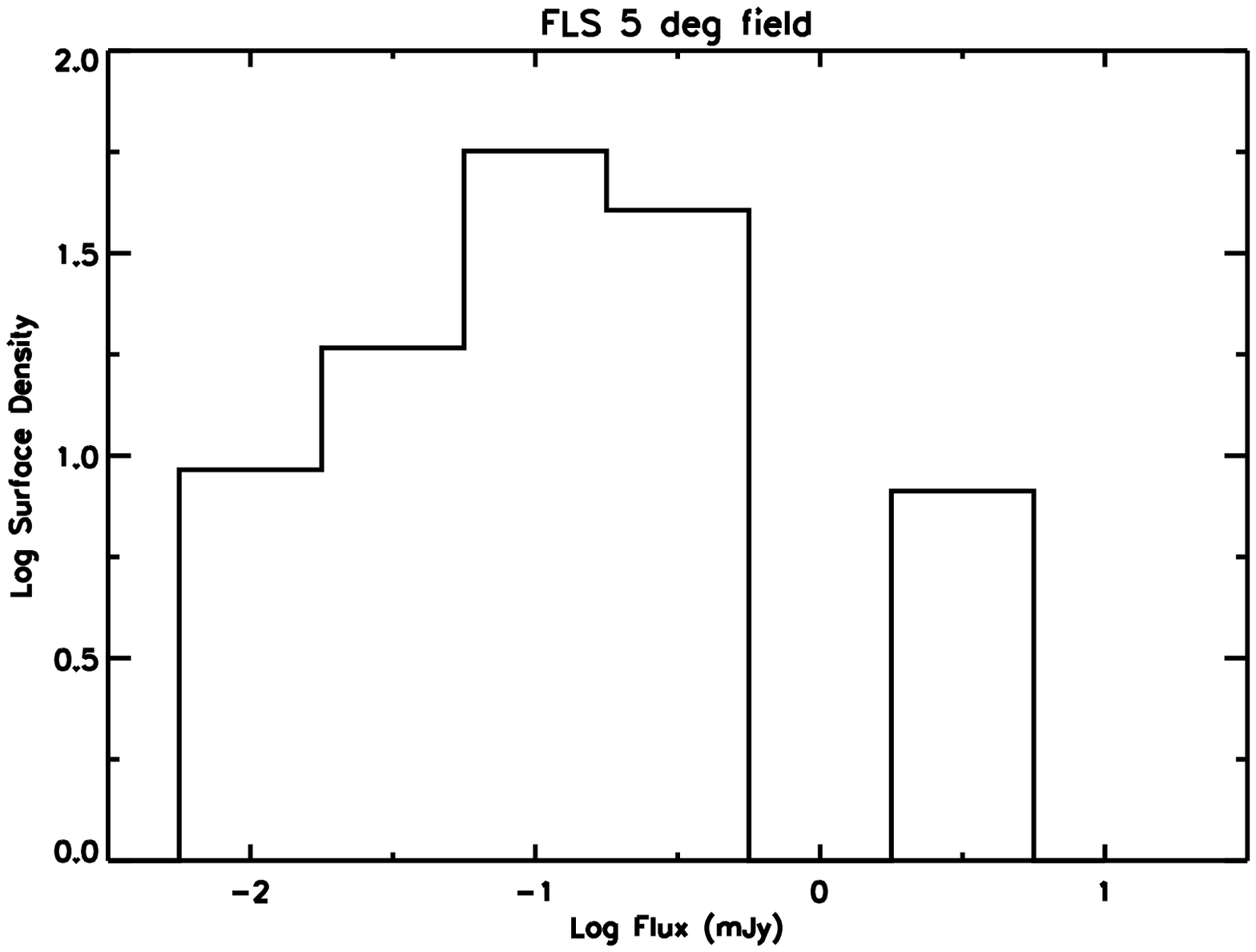}
\includegraphics[angle=0,scale=.55]{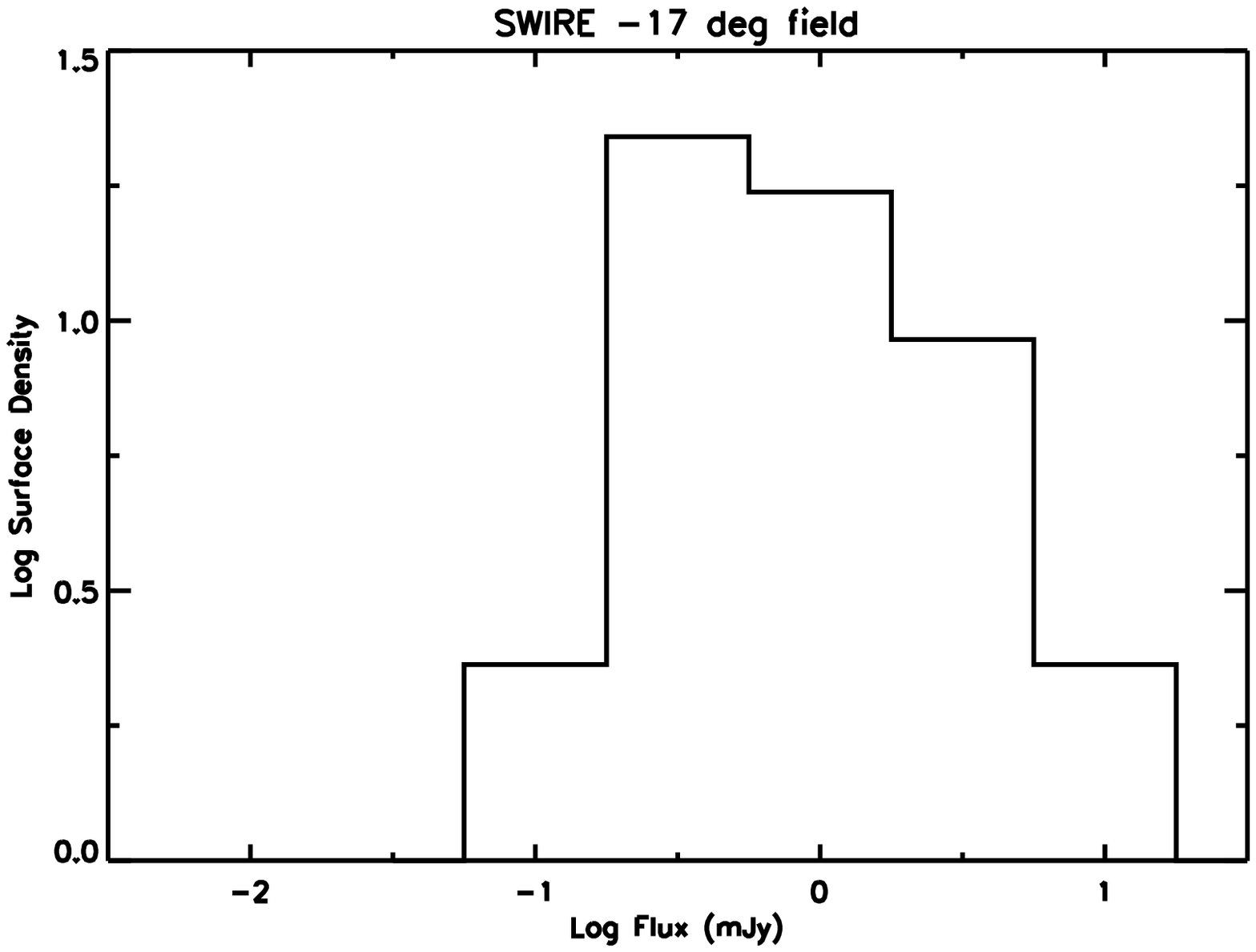}
\end{center}
\vspace{-0.75cm}
\caption{Magnitude histogram distribution for \textit{Spitzer} FLS asteroids 
at $0\degr$ ecliptic latitude (upper), at $5\degr$ ecliptic latitude
(middle), and at $\-17\degr$ ecliptic latitude for the SWIRE field (bottom).
\label{fig:spitzer_histdist_3panel_fig2}}
\end{figure}
\clearpage


\begin{figure}
\plotone{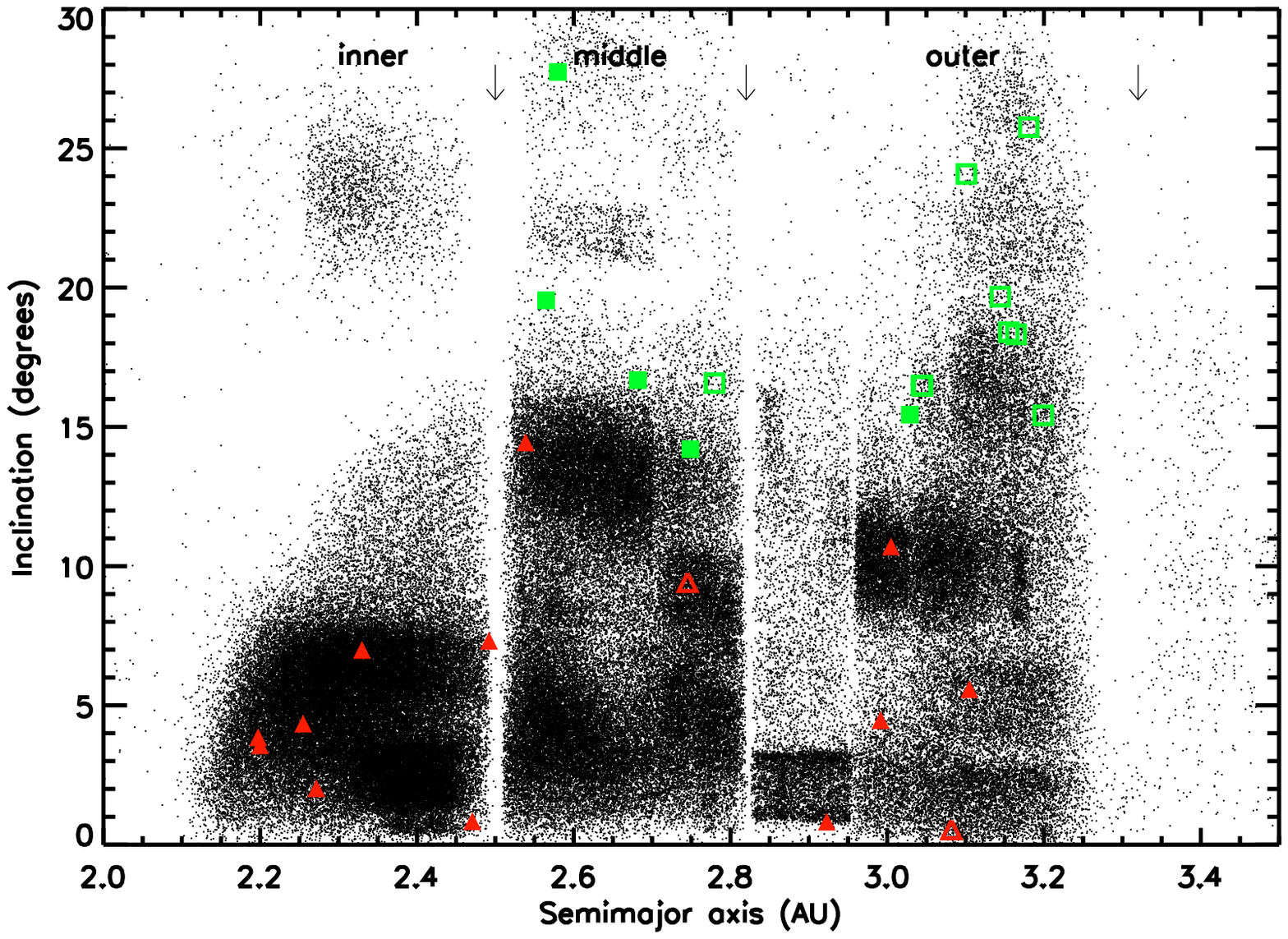}
\caption{Proper orbital elements {\it a}\ (orbital semi-major axis) vs.
{\it i}\ (orbital inclination) of asteroids. The \textit{Spitzer} SWIRE 
and FLS asteroids are denoted respectively by green squares and red 
triangles, while 207,942 numbered asteroids with 
orbital elements in the ASTORB files are plotted with black dots. Open
colored symbols indicate asteroids with derived albedos, $p_{V}$
less than or equal to 0.1, while filled symbols denote asteroids 
with $p_{V} > 0.1$. The location of the major Kirkwood gaps as defined by 
\citet{parker08} are indicated by the labels and vertical arrows.
\label{fig:sdssmoc4_fig3}}
\end{figure}
\clearpage


\begin{figure}
\epsscale{1.0}
\plotone{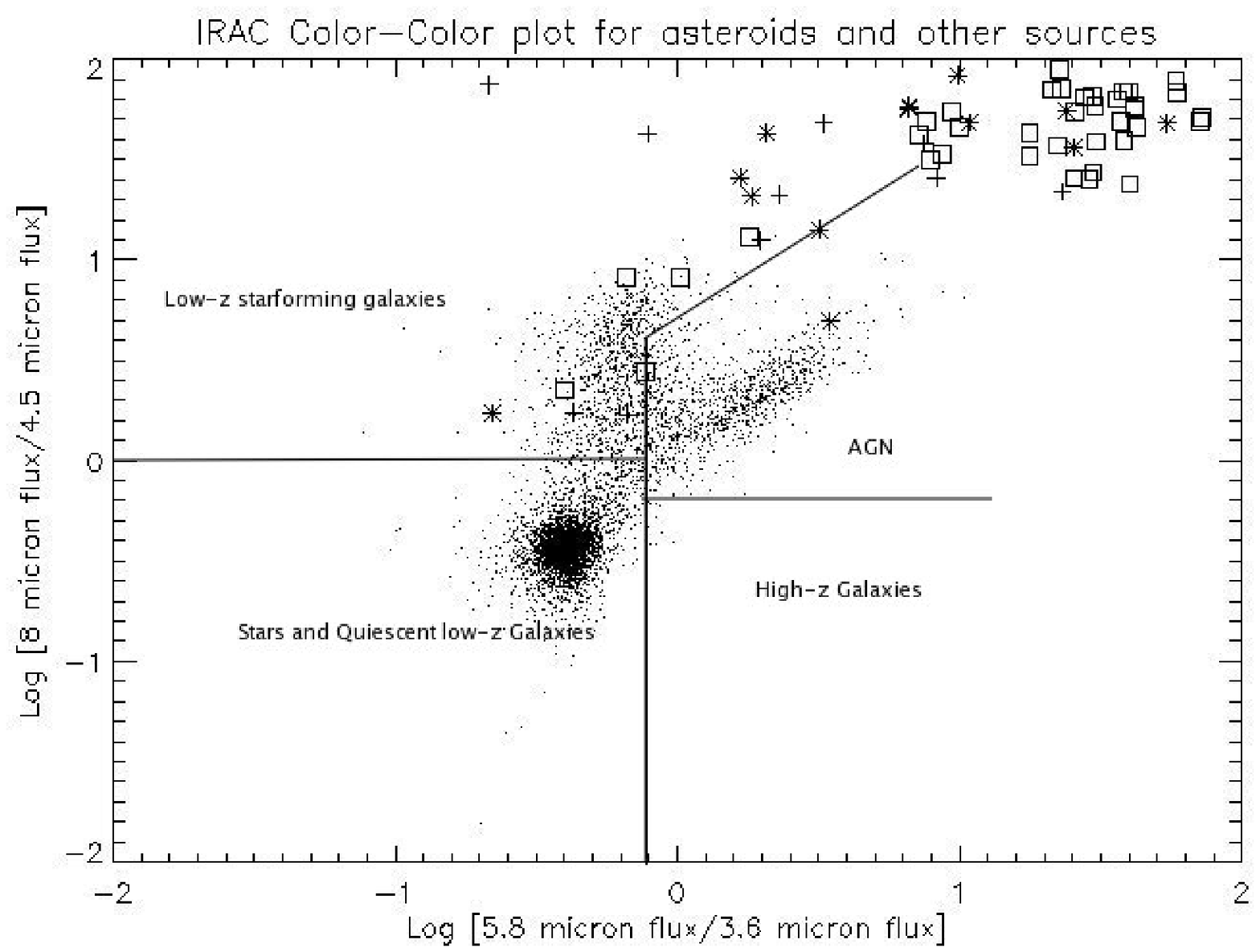}
\caption{IRAC colors of astronomical objects and asteroids. The plus 
signs correspond to asteroids in the FLS 5$\degr$ field, asterisks 
correspond to asteroids in the FLS 0$\degr$ field and boxes 
correspond to asteroids in the SWIRE field. This plot is an adaptation of 
\citet{lacy04} as modified by \citet{elr06}.
\label{fig:color_fig4}}
\end{figure}
\clearpage

\end{document}